\documentclass[sigconf]{acmart}

\usepackage{xcolor}
\usepackage{graphicx}
\usepackage{tikz}
\usepackage{tikzsymbols}
\usetikzlibrary{positioning, shapes.geometric, arrows,patterns,decorations.markings,pgfplots.groupplots}
\usepackage{pgfplots,pgfplotstable} 
    \usepgfplotslibrary{statistics}

\usepackage{algorithm}
\usepackage{algpseudocode} 
\usepackage{amsthm}

\usepackage{tablefootnote}

\usepackage{multirow}
\usepackage{makecell}
\usepackage{enumitem}
\usepackage{amsmath}
\usepackage{graphicx}
\usepackage{algorithm}
\usepackage{algpseudocode}
\usepackage{color}
\usepackage[]{hyperref}
\usepackage[]{caption}
\usepackage{subcaption}
\usepackage{pifont}

\pgfmathdeclarefunction{gauss}{2}{%
  \pgfmathparse{1/(#2*sqrt(2*pi))*exp(-((x-#1)^2)/(2*#2^2))}%
}

\usepackage{circuitikz}
\usetikzlibrary{spy, calc}

\usepackage{adjustbox}

\definecolor{mycolor1}{RGB}{251,180,174}
\definecolor{mycolor2}{RGB}{179,205,227}
\definecolor{mycolor3}{RGB}{204,235,197}
\definecolor{mycolor4}{RGB}{222,203,228}
\definecolor{mycolor5}{RGB}{254,217,166}
\definecolor{mycolor6}{RGB}{255,255,204}
\definecolor{wildstrawberry}{rgb}{1.0, 0.26, 0.64}

\pgfplotsset{compat=1.18}

\AtBeginDocument{%
  }

\setcopyright{acmlicensed}
\copyrightyear{2024}
\acmYear{2024}
\acmDOI{XXXXXXX.XXXXXXX}

\acmConference[ASIA CCS ’24]{ACM ASIA
Conference on Computer and Communications Security}{July 01--05,
  2024}{Singapore}
\acmISBN{978-1-4503-XXXX-X/18/06}

\usepackage{orcidlink}

\begin{document}

\title{REPQC: Reverse Engineering and Backdooring Hardware Accelerators for  Post-quantum Cryptography}



 \author{Samuel Pagliarini}
 \orcid{0000-0002-5294-0606}
 \affiliation{%
   \institution{Carnegie Mellon University}
   \city{Pittsburgh}
   \country{USA}
   }
\additionalaffiliation{%
 \institution{Also with Tallinn University of Technology}
 \city{Tallinn}
 \country{Estonia}
 }

 \email{pagliarini@cmu.edu}

 \author{Aikata Aikata}
 \orcid{0000-0003-0934-2982}
 \affiliation{%
   \institution{Graz University of Technology}
   \city{Graz}
   \country{Austria}}
 \email{aikata@iaik.tugraz.at}

 \author{Malik Imran}
 \orcid{0000-0002-1900-6387}
 \affiliation{%
   \institution{Tallinn University of Technology}
   \city{Tallinn}
   \country{Estonia}}
   \additionalaffiliation{
 \institution{Malik Imran is now with Queen's University Belfast}
 \city{Belfast}
 \country{UK}
 }
 \email{malik.imran@taltech.ee}

 \author{Sujoy Sinha Roy}
 \orcid{0000-0002-9805-5389}
 \affiliation{%
   \institution{Graz University of Technology}
   \city{Graz}
   \country{Austria}}
 \email{sujoy.sinharoy@iaik.tugraz.at}


\begin{abstract}
  Significant research efforts have been dedicated to designing cryptographic algorithms that are quantum-resistant. The motivation is clear: robust quantum computers, once available, will render current cryptographic standards vulnerable. Thus, we need new Post-Quantum Cryptography (PQC) algorithms, and, due to the inherent complexity of such algorithms, there is also a demand to accelerate them in hardware. In this paper, we show that PQC hardware accelerators can be backdoored by two different adversaries located in the chip supply chain. We propose REPQC, a sophisticated reverse engineering algorithm that can be employed to confidently identify hashing operations (i.e., Keccak) within the PQC accelerator -- the location of which serves as an anchor for finding secret information to be leaked. Armed with REPQC, an adversary proceeds to insert malicious logic in the form of a stealthy Hardware Trojan Horse (HTH). Using Dilithium as a study case, our results demonstrate that HTHs that increase the accelerator's layout density by as little as 0.1\% can be inserted without any impact on the performance of the circuit and with a marginal increase in power consumption. An essential aspect is that the entire reverse engineering in REPQC is automated, and so is the HTH insertion that follows it, empowering adversaries to explore multiple HTH designs and identify the most suitable one.
\end{abstract}





\maketitle

    \section{Introduction}\label{sec:intro}
The ability to communicate with people worldwide is the cornerstone of the modern globalized society. It also, in turn, exposes several attack surfaces for security and privacy breach. Public-key cryptographic schemes like RSA and ECC have been protecting us against them for the past two decades. They allow two parties who wish to communicate to start a key exchange over an insecure channel and later switch to a simpler private-key cryptographic scheme. Consequently, several networking protocols, like the Transport Layer Security (TLS) protocol, heavily depend on these schemes.

Despite their current effectiveness, the impending dawn of quantum computers threatens the security of classical cryptographic algorithms. Recognizing this challenge, in 2017, the National Institute of Standards and Technology (NIST)~\cite{nist_pqc} initiated a call for post-quantum Key Encapsulation Mechanisms (KEM) and Digital Signature Algorithms (DSA) to safeguard communications and data exchange in a post-quantum scenario. In early 2023, NIST selected several schemes for standardization, including the lattice-based candidates Kyber~\cite{kyber_nist_round3} and Dilithium~\cite{dilithium_nist_round3}. While the standards are currently being drafted, it is important to emphasize that many federal agencies already have mandates for adopting them \cite{memo}.

PQC algorithms will be gradually deployed in various application domains, and computed on a wide range of software and hardware platforms. While software implementations of PQC offer flexibility and ease of implementation, dedicated hardware platforms like a Field-Programmable Gate Array (FPGA) or an Application Specific Integrated Circuit (ASIC) offer significant speedups\footnote{This was true for classical cryptography and remains true for PQC.}. FPGAs do offer flexibility since they can be reprogrammed, whereas ASICs deliver significantly higher performance than FPGAs but are static in nature and cannot be reprogrammed. For example, the FPGA implementation reported in \cite{kali} outperforms the Cortex-M4 one reported in \cite{BotrosKS19} by 363$\times$. 
Various FPGA-based accelerators are described in~\cite{DBLP:journals/tches/RoyB20, zhou_dil_21,ricci_dil_21, DBLP:journals/iacr/LandSG21,gaj_dil_21, HuangHLW20, XingL21, DangFAMNG20, iacr/NiasarAK21, NiasarAK21, FritzmannSS20, AlkimELNP20, BotrosKS19, DangMG21, NiKKOL22, kali, gmu, sabdil}. Similarly, hardware accelerators tailored for the ASIC platform are reported in~\cite{BanerjeeUC19,FritzmannSS20,XinHYZYCZ20,NiasarAK21,tcas1_ZhaoXXH22,ashes, saber_jcen, saber_tcasII}. For the time being, while the transition to PQC is taking place, FPGAs are suitable platforms. Once standards are finalized and widely adopted, ASIC designs will be more advantageous since they display smaller area footprint, low power, and higher performance. 

For the secure implementation of PQC schemes, we must consider several attack scenarios. Flaws in the protocols or underlying mathematical assumptions can render a cryptographic scheme \textit{weak}. For example,~\cite {sike_attack} reports an attack on an isogeny-based PQC scheme, leading to its exclusion from NIST's consideration.  
Assuming the schemes have withstood scrutiny and are mathematically sound, their hardware implementations may still be vulnerable to side-channel analysis (e.g., SPA, DPA) or fault attacks. These attacks vary in the level of intrusiveness -- in principle, side-channel attacks can be performed even remotely \cite{remote_fpga} whereas fault attacks require physical access to the device under attack. The Deep Learning attack on Kyber~\cite{kyb_attack} is an example of a side-channel attack on an otherwise mathematically sound scheme that succeeds, as expected. 

There are several known techniques to protect against side-channel and fault attacks \cite{reparez_masking2015, OSPG_2018, maskedSaberSW, abubakr21lightweight, dilFritzmannBRKSVS22,kavach}. These techniques were developed for classical cryptography but, in general, can be applied to PQC hardware as well. 
However, one key assumption remains: the hardware accelerator is assumed to be conceived in a trusted supply chain. This paper assumes an opposing view: \textbf{the supply chain is vulnerable} and may contain rogue elements interested in compromising cryptographic chips. In particular, PQC-capable chips are vulnerable as well. Our study focuses on accelerators intended for ASIC platforms since those have stringent security requirements while also having ambitious power/timing/area requirements. To be precise, the threat we are considering in this paper is that of a \textbf{Hardware Trojan Horse (HTH) inserted into an ASIC PQC accelerator}. We will show that reverse engineering such accelerators can be done by locating their hashing building block, i.e., Keccak.

From the point of view of an adversary attempting to mount an HTH attack, many challenges appear \cite{taxonomy}. In order to be successful, the adversary must decide the HTH's location (`where') and the HTH's insertion form (`how'), concepts which we term \emph{node localization} and \emph{HTH insertion}, respectively. \emph{Node localization} involves determining the optimal insertion point knowing that the adversary does not necessarily enjoy a full understanding of the design. Having to mount an attack into a sea of unnamed logic gates is far from trivial. For \emph{HTH insertion}, on the other hand, the adversary has to play a difficult balancing act between detection and attack success: the more logic the HTH design requires, the more sophisticated the attack may become. However, the larger the footprint of the HTH is, the more susceptible to detection it will become. Additionally, a large HTH is harder to fit into the original design without altering its characteristics. 

\emph{Node localization} can be achieved via reverse engineering approaches that group and classify gates. If successful in his/her reverse engineering effort, the adversary can mount very precise and effective attacks. As for \emph{HTH insertion}, automation is key. However, we highlight that automated insertion of HTHs is an area of active study for which there are no general solutions, neither for classical cryptography nor for PQC. We will show that the complexity of PQC hardware and the unique attributes of one of its building blocks, Keccak, allow us to make clever observations leading to a feasible attack strategy. In other words, the main idea put forward in this work is that specialized hardware for PQC is \textbf{not immune} to many supply chain-related threats, including reverse engineering and subsequent HTH insertion.

\subsection{Related Works on HTHs}
The study of compromising cryptographic primitives through HTHs has garnered considerable attention among researchers. This active line of research is adversarial in nature, where security experts mount attacks to comprehend the real capabilities of an adversary. Ravi \emph{et al.} \cite{first_pqc_trojan} were among the first to propose malicious logic that targets PQC hardware. However, their threat model considers only the Third-Party Intellectual Property (3PIP) setting, where a PQC hardware IP is bought from a malicious third party and comes with a backdoor already inside. Their attack was mounted on FPGA-based accelerators for Kyber and Saber.

In \cite{trojan_red_blue} and \cite{trojan_detect}, the authors used a notion of a blue-team versus red-team for HTH detection, where the red-team fabricated a compromised ASIC, and the blue-team was responsible for analyzing it and finding the HTH~\cite{trojan_detect}. In \cite{trojan_red_blue}, this existence of an HTH is \emph{emulated}. These two works enormously contribute to the field, providing a solid foundation for detecting HTHs through physical inspection. In \cite{sca_trojan}, Ghandali \emph{et al.} devised an HTH that makes a masked hardware implementation of a symmetric-key scheme, PRESENT \cite{present}, behave as if it is not masked by violating timing margins almost on demand. 

Hepp \emph{et al.}~\cite{trojan_blind} proposed a generic methodology for HTH insertion based on the relative `importance' of ASIC gates, leveraging reverse engineering tools. In \cite{trojan_riscv_pqc}, Hepp \emph{et al.} described four different means to compromise a PQC hardware accelerator. Two of those require malicious software to trigger the HTH. All four Trojans are inserted manually. Perez \emph{et al.} \cite{iscas_trojan,trojan_tapeout}, described how to leverage an existing feature in chip design tools called the Engineering Change Order (ECO) flow, for HTH insertion. Remarkably, the authors of~\cite{trojan_riscv_pqc,iscas_trojan,trojan_tapeout} also validated their results on an HTH-compromised fabricated chip. A summary of the many characteristics of related works is given in Tab.~\ref{tab:state}. Notice how automation is rarely present. Furthermore, the few works that display automated HTH insertion are very recent developments.

\begin{table}
\small
  \centering
  \caption{Comparison of related works.}\vspace{-10pt}
\label{tab:state}

  \begin{tabular}{c | c | c | c}
  \hline
\textbf{\multirow{2}{*}{Reference}}                       & \textbf{\multirow{1}{*}{Node}} & \textbf{\multirow{1}{*}{HTH}} & \textbf{\multirow{1}{*}{PQC}}\\ 
{\textbf{}} & {\textbf{localization}} & {\textbf{insertion}} & {\textbf{specific}} \\ \hline 

\cite{first_pqc_trojan} *               &  Manual                    & Manual                 & \ding{51} \\          
\cite{trojan_red_blue,trojan_detect}    &  Manual                    & Manual                & \ding{53} \\          
\cite{sca_trojan}                       &  Manual                    & Manual                 & \ding{53} \\          
\cite{trojan_blind}                     &  Automated                  & Automated                & \ding{53} \\          
\cite{trojan_riscv_pqc} $\dagger$       &  Manual                    & Manual                 & \ding{51} \\          
\cite{iscas_trojan,trojan_tapeout}      &  Manual                    & Automated                & \ding{53} \\          
\textbf{This work}                      &  \textbf{Automated}                   & \textbf{Automated}                & \ding{51} \\          \hline
\multicolumn{4}{l}{\small * Assumes 3PIP setting. $\dagger$ Assumes a SW/HW attack.} \\
\end{tabular}

\end{table}

\subsection{Novelty and Contributions}
\noindent \underline{\textbf{Novelty.}} This is the \textbf{first} paper to propose an automated approach to backdoor PQC hardware accelerators, including the steps of node localization and HTH insertion. 


Being so, this work opens up new perspectives for reverse engineering and backdooring PQC hardware implementations. To this end, our contributions are as follows:
\begin{itemize}[leftmargin=*]
    \item \textbf{Contrast between different attackers concerning design and fabrication-time adversaries.} We provide a contrast between different attacks in the supply chain and how they can compromise a chip. For this, we consider a design-time adversary \texttt{A-IP} and a fabrication-time adversary \texttt{A-FO}, both capable of backdooring a PQC accelerator (see Section~\ref{sec:threat_model}).
    
    \item \textbf{First automated method for PQC backdooring.} We propose the first approach for automated node localization and HTH insertion in PQC hardware. It consists of a tool termed \textbf{REPQC} and a specific \textbf{HTH architecture}. REPQC is an open-source tool we propose for reverse engineering ASIC PQC hardware accelerators. Moreover, we offer a novel HTH architecture that, when mounted on a Dilithium implementation, gives the adversary a message-controlled trigger and leaks the long-term key of the digital signature scheme (see Section~\ref{sec:re}).

    \item \textbf{Descriptive result evaluations.} We support our work with results and layouts of several HTH-compromised 28nm ASIC implementations of hardware accelerators for Dilithium (see Section~\ref{sec:results}).

\end{itemize}

    \section{Background}
\label{sec:background}

In this section, we briefly describe all the elements involved in our attack, including a brief overview of PQC and the target algorithm of our attack -- the NIST-selected DSA scheme Crystals-Dilithium~\cite{dilithium_nist_round3}. Within this signature scheme, we target the hashing of secret values performed using Keccak. Our background explanations also cover the basics of reverse engineering and HTHs.

\subsection{Post-quantum Public-key Cryptography}
KEM and DSA schemes help two parties to initiate a key-exchange; this shared key is then utilized to encrypt/decrypt communication packages. A KEM helps to encapsulate or decapsulate a shared key and a DSA scheme allows senders to sign their encapsulated packages for authenticity. Public-key schemes are also often called asymmetric schemes because the encapsulation/decapsulation or sign/verify keys are distinct. These asymmetric schemes typically have very large key and ciphertext components with respect to their symmetric counterparts. Hence, this exchange is expensive in terms of time and memory and is only done once at the beginning of a communication. 

The KEM-related secret key can be refreshed as and when required. On the other hand, the DSA-related secret and public key pairs are long-term and require validation from a certifying authority every time they are changed. If an attacker gets ahold of a user's digital signature secret key, they can pretend to be the user (i.e.,  impersonation). It can also lead to other critical attack scenarios like man-in-the-middle attacks, where the attacker can conveniently listen to a conversation and gain critical information. Thus, it is crucial that the authentication is protected and does not leak any information.

In the current classical scenario, standardized schemes based on RSA and ECC provide this functionality. These schemes are based on specific computationally hard problems. However, these problems can be solved with ease on a quantum computer\footnote{Technically, a cryptanalytically relevant quantum computer capable of executing Peter Shor's algorithm is required.}. Although quantum computers still need to be stronger to break the classical schemes, experts believe it is only a matter of time. Realizing this, NIST launched a standardization call for secure KEM and DSA in a post-quantum scenario. The PQC NIST competition, as it became known, culminated in 2023 with Kyber~\cite{kyber_nist_round3}, Dilithium~\cite{dilithium_nist_round3}, Falcon~\cite{falcon_nist_round3}, and SPHINCS+~\cite{SPHINCS} declared as winners. Out of these, lattice-based schemes Kyber and Dilithium are considered most compatible for applications that require both functionalities, like the TLS protocol. In this work, we recover the secret key of Dilithium. Hence, we present a brief description of this scheme.

\noindent \underline{\textbf{Dilithium.}}
The computational hardness of Dilithium relies on the Module Learning With Errors and Module Short Integer Solution problems. It has three basic functionalities: key generation, signing, and verification. From an attacker's perspective, the most worthwhile operation to compromise is the signing operation. This is because key generation does not happen frequently and offers less opportunity for attack. On the other hand, the verification procedure requires only the public key, thus presenting zero opportunity for a key-recovery attack. Signing is the only operation that happens frequently and involves the secret key. For more details, please refer to \cite{dilithium_nist_round3}.

\begin{algorithm}[t]
\renewcommand{\algorithmicrequire}{\textbf{Input:}}
\renewcommand{\algorithmicensure}{\textbf{Output:}}
\caption{Signing of Dilithium~\cite{dilithium_nist_round3}, simplified}
\begin{algorithmic}[1]
\Require{Secret-key $sk=\{\rho,K,tr,s_1,s_2,\mathbf{t}_0\}$, and msg $M$}
\Ensure{The signature $(z,h,\widetilde{c})$}
\State $A = \text{ExpandA}(\rho)$
\State $\kappa = 0, (z,h) =\ \perp$
\State \color{red} $\mu = \text{Hash}(tr||M)$ \color{black}\Comment{ \texttt{M will serve as trigger}}
\State \color{red} $\rho' = \text{Hash} (K||\mu)$ \color{black} \Comment{ \texttt{ K is to be leaked}}
\While{$(z,h) = \perp$}
    \State $y = \text{ExpandMask}(\rho',\kappa)$
    \State $w = A\cdot y$
    \color{black} \State $w_1 = \text{HighBits}(\text{$w$},2\gamma_2)$
    \State $\widetilde{c} = \text{Hash}(\mu,w_1), c = \text{SampleInBall($\widetilde{c}$)}$
    \State $z = y + c\cdot s_1$
    \State $r_0 = \text{LowBits($w-cs_2$),$2\gamma_2$} $
    \If{$\parallel z \parallel_{\infty} \geq \gamma_1 - \beta$ or $\parallel r_0 \parallel_{\infty} \geq \gamma_2 - \beta$}\color{black}
        \State $(z,h) = \perp$
    \State \vdots ~ \Comment{\texttt{Omitted operations}}
    \EndIf
\EndWhile
\State \Return $(z,h,c)$

\end{algorithmic}
\label{alg:Dil_sign}
\end{algorithm}

A simplified signing procedure is described in Alg.~\ref{alg:Dil_sign}. The secret consists of various components, including `$K$', which is used to generate a pseudo-random string essential for the subsequent generation of polynomial `$y$' (Steps 4 and 6). Then, we compute `$z$' (Step 10), which serves as the output—an LWE sample whose security relies on the secrecy of `$s_1$' and `$y$'. However, it is crucial to note that knowledge of $y$ alone can potentially lead to the disclosure of information about the secret polynomial $s_1$, making it susceptible to signature forgery. The authors in \cite{s1s2} show that knowledge of $s_1$ is sufficient to break the security of the scheme. Consequently, this presents an interesting attack surface. To obtain $y$, an attacker would only need to know $K$, which undergoes processing by the Keccak hash function (Step 4). This will allow the attacker to know $\rho'$ and consequently $y$. Using this, the attacker can recover $s_1$ from Step 10. This observation leads us to the possibility of devising a clever HTH that uses `$M$' as a trigger to leak `$K$' -- the details are given in Section~\ref{subsec:hth}.

 \noindent \underline{\textbf{Keccak.}}
 Keccak~\cite{Keccak11} is the core of NIST-standardized~\cite{fips_202} Secure Hash Algorithm-3 (SHA-3). Keccak is a ``sponge'' function, which absorbs the input data and then ``squeezes'' a digest. It can also be used as an Expandable Output Function (SHAKE), where it absorbs a small seed and produces a long string of pseudo-random bits. On the other hand, in the hash function mode (SHA-3), Keccak absorbs a large input data and then outputs a brief digest (hash output). For Dilithium, Keccak is employed in four modes: SHA-3/256, SHA-3/512 variants for hashing; and SHAKE128, SHAKE256 for pseudo-random number generation.

\begin{figure}
    \centering
    \begin{tikzpicture}[scale=0.7]
    \foreach \x in{0,...,5}
    {   

        \draw [black] (0,\x ,5) -- (5,\x ,5);
        \draw [black] (\x ,0,5) -- (\x ,5,5);
        \draw [dotted] (5,\x ,5) -- (5,\x ,0);
        \draw [dotted] (\x ,5,5) -- (\x ,5,0);
        \draw [dotted] (5,0,\x ) -- (5,5,\x );
        \draw [dotted] (0,5,\x ) -- (5,5,\x );
    }

    \foreach \x in{0}{  
    \foreach \y in{0}{
        \node[font=\tiny] at (\x-1.45,2.5-\y) {$y_2,x_3$};
    }}

    \foreach \x in{0}{  
    \foreach \y in{1}{
        \node[font=\tiny] at (\x-1.45,2.5-\y) {$y_1,x_3$};
    }}

    \foreach \x in{0}{  
    \foreach \y in{2}{
        \node[font=\tiny] at (\x-1.45,2.5-\y) {$y_0,x_3$};
    }}

    \foreach \x in{0}{  
    \foreach \y in{3}{
        \node[font=\tiny] at (\x-1.45,2.5-\y) {$y_4,x_3$};
    }}

    \foreach \x in{0}{  
    \foreach \y in{4}{
        \node[font=\tiny] at (\x-1.45,2.5-\y) {$y_3,x_3$};
    }}

   \foreach \x in{1}{  
    \foreach \y in{0}{
        \node[font=\tiny] at (\x-1.45,2.5-\y) {$y_2,x_4$};
    }}

    \foreach \x in{1}{  
    \foreach \y in{1}{
        \node[font=\tiny] at (\x-1.45,2.5-\y) {$y_1,x_4$};
    }}

    \foreach \x in{1}{  
    \foreach \y in{2}{
        \node[font=\tiny] at (\x-1.45,2.5-\y) {$y_0,x_4$};
    }}

    \foreach \x in{1}{  
    \foreach \y in{3}{
        \node[font=\tiny] at (\x-1.45,2.5-\y) {$y_4,x_4$};
    }}

    \foreach \x in{1}{  
    \foreach \y in{4}{
        \node[font=\tiny] at (\x-1.45,2.5-\y) {$y_3,x_4$};
    }}
    
    \foreach \x in{2}{  
    \foreach \y in{0}{
        \node[font=\tiny] at (\x-1.45,2.5-\y) {$y_2,x_0$};
    }}

    \foreach \x in{2}{  
    \foreach \y in{1}{
        \node[font=\tiny] at (\x-1.45,2.5-\y) {$y_1,x_0$};
    }}

    \foreach \x in{2}{  
    \foreach \y in{2}{
        \node[font=\tiny] at (\x-1.45,2.5-\y) {\textcolor{blue}{\textbf{$y_0,x_0$}}};
    }}

    \foreach \x in{2}{  
    \foreach \y in{3}{
        \node[font=\tiny] at (\x-1.45,2.5-\y) {$y_4,x_0$};
    }}

    \foreach \x in{2}{  
    \foreach \y in{4}{
        \node[font=\tiny] at (\x-1.45,2.5-\y) {$y_3,x_0$};
    }}

    \foreach \x in{3}{  
    \foreach \y in{0}{
        \node[font=\tiny] at (\x-1.45,2.5-\y) {$y_2,x_1$};
    }}

    \foreach \x in{3}{  
    \foreach \y in{1}{
        \node[font=\tiny] at (\x-1.45,2.5-\y) {$y_1,x_1$};
    }}

    \foreach \x in{3}{  
    \foreach \y in{2}{
        \node[font=\tiny] at (\x-1.45,2.5-\y) {$y_0,x_1$};
    }}

    \foreach \x in{3}{  
    \foreach \y in{3}{
        \node[font=\tiny] at (\x-1.45,2.5-\y) {$y_4,x_1$};
    }}

    \foreach \x in{3}{  
    \foreach \y in{4}{
        \node[font=\tiny] at (\x-1.45,2.5-\y) {$y_3,x_1$};
    }}

    \foreach \x in{4}{  
    \foreach \y in{0}{
        \node[font=\tiny] at (\x-1.45,2.5-\y) {$y_2,x_2$};
    }}

    \foreach \x in{4}{  
    \foreach \y in{1}{
        \node[font=\tiny] at (\x-1.45,2.5-\y) {$y_1,x_2$};
    }}

    \foreach \x in{4}{  
    \foreach \y in{2}{
        \node[font=\tiny] at (\x-1.45,2.5-\y) {$y_0,x_2$};
    }}

    \foreach \x in{4}{  
    \foreach \y in{3}{
        \node[font=\tiny] at (\x-1.45,2.5-\y) {$y_4,x_2$};
    }}

    \foreach \x in{4}{  
    \foreach \y in{4}{
        \node[font=\tiny] at (\x-1.45,2.5-\y) {$y_3,x_2$};
    }}

    \foreach \x in{0}{  
    \foreach \y in{6}{
        \node[font=\tiny] at (\x-1.45,3.85-\y) {3};
    }}

    \foreach \x in{1}{  
    \foreach \y in{6}{
        \node[font=\tiny] at (\x-1.45,3.85-\y) {4};
    }}

    \foreach \x in{2}{  
    \foreach \y in{6}{
        \node[font=\tiny] at (\x-1.45,3.85-\y) {0};
    }}

    \foreach \x in{3}{  
    \foreach \y in{6}{
        \node[font=\tiny] at (\x-1.45,3.85-\y) {1};
    }}

    \foreach \x in{4}{  
    \foreach \y in{6}{
        \node[font=\tiny] at (\x-1.45,3.85-\y) {2};
    }}

    \foreach \x in{0}{  
    \foreach \y in{0}{
        \node[font=\tiny] at (\x-2.15,2.5-\y) {2};
    }}

    \foreach \x in{0}{  
    \foreach \y in{1}{
        \node[font=\tiny] at (\x-2.15,2.5-\y) {1};
    }}

    \foreach \x in{0}{  
    \foreach \y in{2}{
        \node[font=\tiny] at (\x-2.15,2.5-\y) {0};
    }}

    \foreach \x in{0}{  
    \foreach \y in{3}{
        \node[font=\tiny] at (\x-2.15,2.5-\y) {4};
    }}

    \foreach \x in{0}{  
    \foreach \y in{4}{
        \node[font=\tiny] at (\x-2.15,2.5-\y) {3};
    }}

    \foreach \x in{0}{  
    \foreach \y in{0}{
    \foreach \z in{0}{
        \node[font=\tiny] at (\x-1.35,4.0-\y,\z1.5) {0};
        \node[font=\tiny] at (\x-1,4.4-\y,\z1.5) {1};
        \node[font=\tiny,rotate=45] at (\x-0.35,4.9-\y,\z1.5) {$. . .$};
        \node[font=\tiny,rotate=45] at (\x0.3,5.6-\y,\z1.5) {$w-1$};
    }}}

    \foreach \x in{0}{  
    \foreach \y in{7}{
        \draw[black, |-|] (\x-2.5,3.0\y) -- (\x-2.5,5.0-\y);
        \draw[black, |-|] (\x-2.0,-2.5\y) -- (\x3.1,-2.5\y);
        \draw[black, |-|] (\x-2.45,3.45\y) -- (\x-0.45,5.455\y);
    }}

    \foreach \x in{0}{  
    \foreach \y in{7}{
        \node[font=\tiny] at (\x0.6,4.25-\y) {$x$};
        \node[font=\tiny] at (\x-2.675,0.41\y) {$y$};
        \node[font=\tiny] at (\x-1.75,4.41\y) {$z$};
    }}
    
    \end{tikzpicture}\vspace{-10pt}
    \caption{Keccak state as a $5\times5\times64$ 3D matrix, for keccak-f[1600]. Each box in the Keccak state represents one bit.}\label{fig:keccak_state_buffer}
    \end{figure}
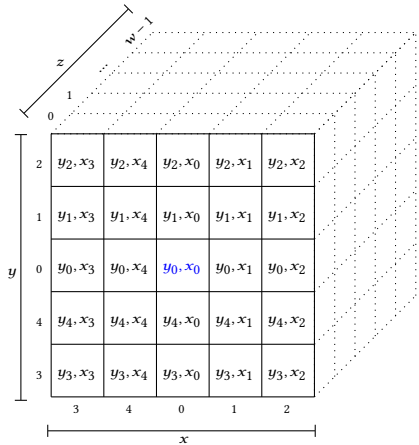

At the core of Keccak is its permutation function denoted as keccak-f[b], where $b=25 \times 2^l$ and $0\leq l \leq 6$. The keccak-f[b] is defined over $s \in \mathcal{Z}_2^b$, where $b$ represents the width of the permutation. In essence, the permutation used in SHA-3 variants can be configured with different parameters, including capacity, rate, etc. For more detailed information, we refer readers to ~\cite{Keccak11, fips_202}. The keccak-f[b] permutation operates on a state $a$ that is defined as a three-dimensional array denoted as $a[5][5][w]$, where $w=2^l$. For the maximum permutation width (keccak-f[1600] and $l=6$), $w$ is equal to $2^6=64$. Consequently, the state $a$ is represented by a three-dimensional array with dimensions of $[5][5][64]$. A visual representation of the state $a$ is provided in Fig.~\ref{fig:keccak_state_buffer}. It is worth mentioning that a single permutation computation in Keccak relies on $12+2 \times l$ rounds, each round comprising a combination of five linear and non-linear steps, namely \{$\theta$, $\rho$, $\pi$, $\chi$, and $\iota$\}~\cite{fips_202}.

\subsection{Reverse Engineering}

Reverse Engineering can be defined as the process of creating a \emph{representation} of a piece of hardware by someone other than the original designer(s). In other words, it is the process of making sense of an Integrated Circuit (IC) without \emph{a priori} knowledge about its implementation. IC reverse engineering has two sides to it: physical and logical. On the physical side, many studies have disclosed how to perform product tear-down, IC sample preparation, delayering\footnote{Delayering or delamination is the process of removing the layers of a chip one at a time. It often is performed with chemical etching.}, imaging\footnote{Imaging processes based on SEM or X-ray are typically employed.}, and stitching\footnote{Stitching is the process of aligning the many individual images.} of the acquired images to reconstruct the layout of a circuit. This process has many challenges \cite{re_challenges}, including the precision of the delamination, the throughput and resolution of the imaging, and the quality of the stitching. 

However, in this work, we are not dealing with these physical requirements since our adversary possesses a perfect representation of the layout. This becomes clear in our threat model in Section~\ref{sec:threat_model}, when we reveal that our adversary is a foundry engineer. In this work, we are dealing with logical reverse engineering: assuming a layout has been acquired and transformed into a netlist (schematic), one still has to comprehend it since the obtained netlist is rather opaque to a human. 
In order to attempt to give meaning to `blind' netlists, different analytical approaches can be applied. A typical first step is netlist partitioning, primarily for the sake of making the problem tractable. Any partitioning solution that aids in the understanding of the design is a good partitioning.
A further approach is the separation of datapaths from control logic. RELIC \cite{relic, neta} is a prime example of a tool for this task. It builds on the observation that wires in the control logic tend to have a non-repeatable \emph{structure}, whereas wires in a datapath will have a structure similar to other wires of the same word. RELIC's approach is not perfect, but it tends to be rather insightful. In Appendix~\ref{app:relic}, we provide examples of distributions that illustrate the output of RELIC's analysis.

Another classic problem in logical reverse engineering is register grouping \cite{re_dana}. The problem can be summarized as follows: for a netlist containing $N$ individual flip-flops, how to best arrange $N$ into registers that have the same properties? For instance, in a circuit that has a 64-bit datapath, one would be inclined to look for 64-bit registers that hold the input data to a 64-bit operation. Register grouping can be performed based on similarity, as discussed in \cite{wordrev}.

Reverse engineering tools can be used for benign and malicious purposes and, in this work, we will use a \emph{cocktail} of these tools for the purpose of \emph{node localization}. Once nodes of interest are found, we then proceed with HTH insertion.

\subsection{Hardware Trojan Horses}

For decades, the IC industry has continuously become more and more distributed and globalized. Consequently, no IC design today is conceived by a single company and many entities are involved in what has been termed the fabless model. In this model, design-related activities are handled by fabless design houses while the fabrication is handled by a third-party silicon foundry\footnote{This is an oversimplification that abstracts away post-fabrication steps.}. The design house is the rightful IP owner but it has no oversight or control claim over the fabrication process. In other words, the fabrication process is considered untrusted by definition, and it is precisely at fabrication time that HTHs can be inserted.

HTHs have been studied for nearly two decades now, which has yielded a well-established taxonomy \cite{taxonomy} and a database \cite{trusthub}. Among the many forms an HTH may take, additive HTHs stand out. Those are pieces of logic added on top of the original circuit design, hence the name. They may leak some secret information directly via a primary output of the chip, or simply cause the chip to stop working altogether, akin to a Denial of Service (DoS) attack. A specialized form of an additive HTH is a \textbf{side-channel HTH} that induces a controlled/modulated amount of power consumption through which information can be leaked. This HTH style has been analytically described in \cite{moles} and demonstrated as silicon-viable in \cite{trojan_tapeout}. By definition, a side-chanel HTH is harder to detect since no primary outputs of the chip are used for information leakage. Hence, we will adopt the side-channel style of HTH due to its proven viability and inherent stealthiness.


Conceptually, HTHs have two components: a trigger and a payload \cite{taxonomy}. The trigger is the logic that determines \emph{when} the malicious activity should start. A `good' trigger is one that is stealthy in the sense that only the adversary can reasonably (and deterministically) activate it. Examples of reasonable triggers are comparators and counters. The payload, on the other hand, is the component that determines \emph{how} the HTH will interfere with the chip's functionality. The most basic form of a payload is an XOR gate that flips some signal of the original design at will, thus corrupting the underlying computation. Ring oscillators \cite{trojan_tapeout}, shift registers \cite{trojan_blind}, and even antennas \cite{trojan_wireless} can be considered as payloads for these HTHs as they can demonstrate modulated power consumption.

A distinction between hardware backdoors and HTHs is not always clear in the literature. A hardware backdoor tends to be inserted by someone at design time leading to compromised code or 3PIP~\cite{first_pqc_trojan}. On the other hand, the term HTH tends to refer to malicious logic inserted at fabrication time, as in \cite{trojan_tapeout}. We will consider both scenarios in our work and refer to them solely as HTH for the sake of simplicity.

    \section{Threat Model}
\label{sec:threat_model}

\begin{figure}
\centering
\includegraphics[width=0.95\linewidth]{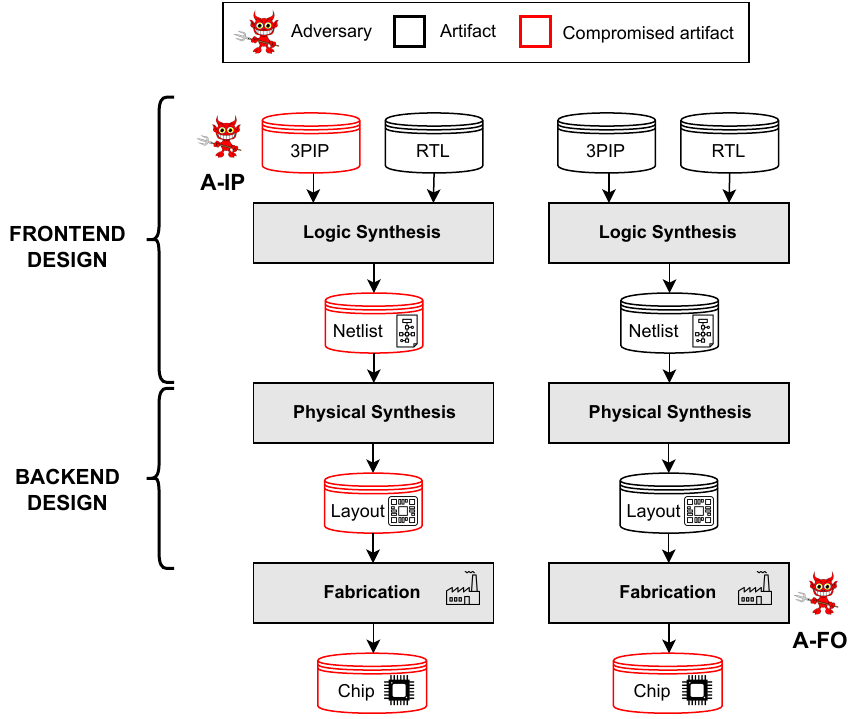}\vspace{-10pt}
\caption{Locations of the adversaries \texttt{A-IP} and \texttt{A-FO} in the supply chain and design flow.}
\label{fig:supply}
\end{figure}

In this work, we consider two adversarial settings, namely the 3PIP and the untrusted foundry settings. We term the adversaries associated with these as \texttt{A-IP} and \texttt{A-FO}, respectively. The presence of these two adversaries in the IC design flow and supply chain is shown in Fig. \ref{fig:supply}. 

The design flow depicted in Fig. \ref{fig:supply} can be understood as a series of transformations. Logical synthesis is the process of translating a human-readable description of a design in hardware description language into a schematic of interconnected logic cells (AND, OR, XOR, flip-flops, etc.). Physical synthesis converts these logically connected cells into a physical representation, that is, a layout of the chip. Finally, fabrication transforms a layout into a physical device. Notice that once an artifact is compromised in this chain of transformations, the subsequent representations remain compromised.

\texttt{A-IP} is an external adversary that has provided a compromised IP to an honest design house. Such an IP is assumed to be a source code described in a hardware description language. This adversary must have a fair understanding of PQC in order to compromise a PQC hardware accelerator. However, he/she does not need to understand reverse engineering techniques or backend design since, to him/her, the design is transparent and human-readable. 

Contrary to this, \texttt{A-FO} is a foundry engineer that has the means to modify the layout before it goes into production. \texttt{A-FO} can also convert the layout to a netlist representation with ease \cite{regds}. Furthermore, \texttt{A-FO} should have: (i) a good understanding of PQC algorithms, (ii) reverse engineering skills, and (iii) advanced chip design skills. For (i), since the PQC algorithms are open and standardized, the adversary may have advanced knowledge about their requirements and what their building blocks are. For (ii), the adversary can adopt REPQC, the contribution of this work. For (iii), we assume the \texttt{A-FO} adversary is as skilled as the rogue engineer described in \cite{trojan_tapeout}, in line with the assumptions made in most of the HTH literature. We further assume \texttt{A-FO} performs HTH insertion via the ECO flow.

On the surface, one might assume that \texttt{A-IP} is a more powerful adversary than \texttt{A-FO}. However, \texttt{A-IP} has a severe disadvantage: the compromised artifact generated by \texttt{A-IP} appears very early in the flow. Being so, many detection approaches are applicable, including verification by simulation, information flow tracking, and taint analysis. None of those approaches apply to \texttt{A-FO}. Only physical inspection of a (suspicious) chip can guarantee detection, but physical inspection of fabricated chips is neither cursory nor cost-efficient.  

Next, we define what the trigger and the payload ought to look like in the HTHs inserted by \texttt{A-IP} and \texttt{A-FO}.


\newtheorem*{remark}{Trigger design}

\theoremstyle{definition}
\newtheorem*{inner}{\innerheader}
\newcommand{\innerheader}{}
\newenvironment{defi}[1]
 {\renewcommand\innerheader{#1}\begin{inner}}
 {\end{inner}}

\begin{defi}{Trigger definition}
We assume the trigger is the message $M$ (or part of it), as highlighted in red in line 3 of Alg. \ref{alg:Dil_sign}. The adversary can select any message $M$ of his/her liking for this purpose. The same assumption is made for \texttt{A-IP} and \texttt{A-FO}. Physically, the trigger logic is a comparator.
\end{defi}

\begin{defi}{Payload definition}
For the payload, we assume both adversaries are going to insert malicious logic that can modulate power consumption on demand in order to enable side-channel leakage. This is achieved by modulating power based on the value of $K$, as highlighted in red in line 4 of Alg. \ref{alg:Dil_sign}. Physically, the payload is a ring oscillator.
\end{defi}

Notice that, although the definitions above imply that the adversaries possess  knowledge about the PQC algorithm, they do not have any \emph{a priori} knowledge of the circuit under attack.


\algnewcommand\algorithmicforeach{\textbf{for each}}
\algdef{S}[FOR]{ForEach}[1]{\algorithmicforeach\ #1\ \algorithmicdo}

    \section{Our Proposed REPQC Tool and HTH Architecture}
\label{sec:re}
Let us first refer again to Fig. \ref{fig:supply} before describing our proposed REPQC tool and HTH architecture. Note that only \texttt{A-FO} has to perform RE since the entire design is transparent to \texttt{A-IP}. \texttt{A-FO}, on the other hand, has to identify where the two elements ($M,~K$) exist in the design blindly, possessing only a netlist where the nodes have no name or any otherwise identifying characteristic. Hence, in this section, we describe an approach to empower the adversary \texttt{A-FO} with as much knowledge about the design as that possessed by \texttt{A-IP}. 

For this, we use a cocktail of RE tools: RELIC $\rightarrow$ REPCA $\rightarrow$ REDPEN $\rightarrow$ REPQC. The first three are from the NETA suite~\cite{neta}, whereas REPQC is the main contribution of this paper. These tools are chained together as described in Fig.~\ref{fig:flowchart}.

\tikzstyle{startstop} = [rectangle, rounded corners, 
minimum width=1cm, 
minimum height=0.5cm,
text centered,
fill=gray!25,
text width=1.5cm, 
draw=black,
font=\footnotesize]

\tikzstyle{process} = [rectangle, 
minimum width=1cm, 
minimum height=0.5cm, 
text centered, 
text width=1.5cm, 
draw=black, 
fill=gray!25,
font=\footnotesize]

\tikzstyle{decision} = [diamond, 
minimum width=1cm, 
minimum height=0.5cm, 
text centered, 
draw=black,
font=\footnotesize]

\tikzstyle{arrow} = [thick,->,>=stealth]

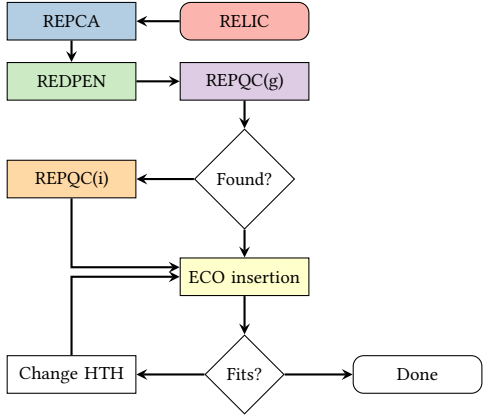
\begin{figure}
\centering
\begin{tikzpicture}[node distance=0.8cm]

\node (pro1) [startstop, fill=mycolor1] {RELIC};
\node (pro2) [process, left of=pro1, xshift=-1.5cm, fill=mycolor2] {REPCA};
\node (pro3) [process, below of=pro2, fill=mycolor3] {REDPEN};
\node (pro4) [process, right of=pro3, xshift=1.5cm, fill=mycolor4] {REPQC(g)};
\node (dec1) [decision, below of=pro4, yshift=-0.5cm] {Found?};

\node (pro5a) [process, below of=dec1, yshift=-0.5cm, fill=mycolor6] {ECO insertion};
\node (pro5b) [process, left of=dec1, xshift=-1.5cm, fill=mycolor5] {REPQC(i)};

\node (dec2) [decision, below of=pro5a, yshift=-0.5cm] {Fits? };

\node (pro6a) [startstop, right of=dec2, xshift=+1.5cm,fill=white] {Done};
\node (pro6b) [process, left of=dec2, xshift=-1.5cm,fill=white] {Change HTH};

\draw [arrow] (pro1) -- (pro2);
\draw [arrow] (pro2) -- (pro3);
\draw [arrow] (pro3) -- (pro4);
\draw [arrow] (pro4) -- (dec1);

\draw [arrow] (dec1) -- (pro5b);
\draw [arrow] (dec1) -- (pro5a);
\begin{scope}[transform canvas={yshift=.4em}]
  \draw [arrow, shorten <=.4em] (pro5b) |- (pro5a);
\end{scope}
\draw [arrow] (pro5a) -- (dec2);
\draw [arrow] (dec2) -- (pro6b);
\draw [arrow] (dec2) -- (pro6a);
\draw [arrow] (pro6b) |- (pro5a);
\end{tikzpicture}
\vspace{-10pt}
\caption{Flowchart of the entire attack.}
\label{fig:flowchart}
\end{figure}

\noindent \underline{\textbf{NETA-suite Reverse Engineering Tools.}}
In Fig.~\ref{fig:flowchart}, RELIC takes as input a design netlist and produces a list of flip-flops ordered by Z-score. The Z-score ($\in [0,\infty)$) is a measure of how much a flip-flop is likely to be part of the control (or data) path of a design. The higher the value, the more likely the flip-flop is unique and, therefore, part of the control logic. The closer the number is to zero, the more likely the flip-flop is part of the datapath. Examples of RELIC's output are in Appendix~\ref{app:relic}.

REPCA takes as input the same netlist and produces a list of groups of flip-flops, which we term registers. If REPCA does a good job, these registers should match the high-level registers in the original design description. 
REDPEN, a much simpler RE tool, takes a netlist as input and identifies all flip-flop dependencies. For example, if at any given clock cycle $c_i$, the content of a flip-flop $f_1$ depends on flip-flop $f_2$ at $c_{i-1}$, there is a dependency $f_2 \rightarrow f_1$.

RELIC and REPCA make use of the concept of features for assigning Z-scores and groupings to individual flip-flops. These features are varied and can be combined for making a recipe. The default recipe in NETA makes use of 20 features together. In our experiments, we use much more modest (and yet sufficient) recipes. Our RELIC recipe has only one feature: FANOUT. Our REPCA recipe has two features: INPUT and OUTPUT. The meaning of these features is explained in detail in the NETA repository \cite{neta_repo}. Our choice of these simple features is discussed in Section~\ref{sec:our_repqc}.

\subsection{Our Proposed Tool: REPQC}\label{sec:our_repqc}

First, let us start with a high-level description of REPQC as given in Alg.~\ref{alg:repqc}. REPQC takes as input three files that correspond to the Z-scores from RELIC, the groups from REPCA, and the dependencies from REDPEN. The Z-scores for each flip-flop are stored in the $FFs$ list, $GROUPs$ list stores the groups, and the map $DEPs[]$ stores dependencies.

\begin{algorithm}[t]
  \caption{REPQC}
  \begin{algorithmic}[1]
  \renewcommand{\algorithmicrequire}{\textbf{Input:}}
\renewcommand{\algorithmicensure}{\textbf{Output:}}

  \Require{$file.scores,file.groups,file.deps$}
    \Ensure{List of flip-flops that carry $M$ and $K$}

        \State $FFs \gets parse(file.scores)$
        \State $GROUPs\gets parse(file.groups)$
        \State $DEPs[]\gets parse(file.deps)$

        \ForEach {$f \in FFs $}
            \If{$f.fanin/out \notin [FIF,FIC]/[FOF,FOC]$} 
                \State $FFs.remove(f)$
            \EndIf
        \EndFor
        \ForEach {$f \in FFs $}
            \ForEach {$reg \in GROUPs $}
                 \ForEach {$member \in reg $}
                 \If{$DEPs$[$f$,$member$]} 
                    \State $member.hit \gets TRUE$
                    \State $reg.hits \gets reg.hits + 1$
                 \EndIf
                 \EndFor
            \EndFor
        \EndFor

         \ForEach {$reg \in GROUPs $} 
            \If{$reg.hits < 64$} 
               \State $GROUPs.remove(reg)$
            \Else
                \ForEach {$member \in reg $}
                    \If{$member.hit == FALSE$} 
                \State $reg.remove(member)$
                \EndIf
            \EndFor
            \EndIf
        \EndFor

        \State $sort(GROUPs)$
        
      \State \Return $GROUPs[0].members $\Comment{ranked by Z-score}
  \end{algorithmic}
  \label{alg:repqc}
\end{algorithm}

The first step is to filter out flip-flops of interest. The goal is to find the flip-flops that implement the Keccak state $a$ USING their fanin and fanout properties. 
For a given flip-flop $f$, its fanin and fanout are the numbers of flip-flops in $FFs$ for which there is a dependency  $f_i \rightarrow f$ and $f \rightarrow f_i$. Note that this is the definition of sequential fanin/fanout.
Next, for every flip-flop in the design, we check whether the fanin and fanout are between the respective floor and ceiling, i.e., $[FIF, FIC],[FOF, FOC]$ (line 5).

\input{keccakplot} 

Reasonable assumptions for these four constants can be determined primarily from analyzing the Keccak specification. The Keccak state, as shown in Fig.~\ref{fig:keccak_state_filled}, is assumed to be implemented as a single register with 1600 flip-flops. Due to the iterative nature of the Keccak rounds, the state updates itself after every iteration by applying the round functions \{$\theta$, $\rho$, $\pi$, $\chi$, and $\iota$\}. For keeping the Keccak hash operation secure, the state cannot be arbitrarily connected to random flip-flops. The connections must be such that the Keccak state can be loaded, processed (permutated), and read.

\begin{figure*}
\centering
\includegraphics[width=0.8\linewidth]{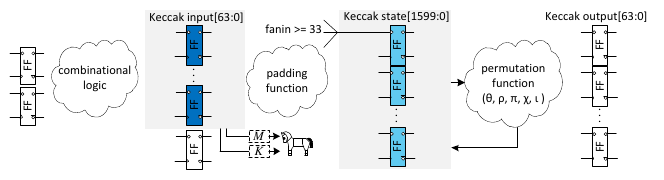}
\vspace{-10pt}
\caption{The Keccak state is composed of 1600 flip-flops that have predictable fanin/fanout properties. The goal is to find a register with 64 flip-flops that is the input to Keccak and therefore holds the message $M$ and the component $K$.}
\label{fig:target}
\end{figure*}

In Fig.~\ref{fig:keccak_state_filled}, we show how bit $a[0][0][0]$ is updated after one round. Notice that the new value of $a[0][0][0]$ depends on exactly 32 other elements of $a$ and itself. Hence, the $FIF$ value is at least 33. 
Similar assumptions can be made for $FOF$, where the permutation logic determines, assuredly, that 34 dependencies exist. The actual fanout number may be higher because the Keccak state flip-flops may be connected to other units of the accelerator. In other words, 
34 can be interpreted as a floor for the actual number of fanouts ($FOF$).

With these assumptions in place, a very large portion of the design flip-flops are immediately rejected as Keccak state candidates. Next, we use the notion of a `hit' to find flip-flops that are inputs to Keccak. In order to identify the Keccak input, we compare every surviving candidate flip-flop against all groups. If a dependency is found, the group member is marked with a hit (line 13) and the number of hits associated with a group $reg$ is incremented by one (line 14). As a reminder, our attack does not aim to compromise Keccak itself. Our goal is to use Keccak as an anchor for RE. This is visualized in Fig.~\ref{fig:target}. It is the input to Keccak that is of interest since it may hold the message $M$ that is the trigger for the HTH. The input also holds the component $K$ that we wish to leak. 

Next, REPQC looks at the groups of interest. A group that has received less than 64 hits is not a reasonable candidate (line 20) because the input to Keccak is, by definition, 64 bits wide. For groups that have more than 64 hits, we have to check where these hits come from. Any member of a group that was never `hit' by a candidate flip-flop is eliminated since it cannot be an input to Keccak (line 25). After all bad candidates are eliminated, the groups are sorted by their Z-score (line 30), and finally, the 64 lowest-scoring elements of the lowest-scoring group are returned (line 31). For more details and optimizations, we refer the readers to our repository  \cite{OURREPO}.

Fig.~\ref{fig:target} shows the high-level objective of REPQC: finding the dark blue flip-flops based on the properties of the light blue flip-flops.


\noindent \underline{\textbf{RELIC Features utilized in REPQC.}} For assigning Z-scores to flip-flops, the only feature we use is FANOUT. Meaning that flip-flops are compared against others based on their fanouts, and this determines whether a flip-flop is more data- or control-oriented. The insight here is that the permutation functions create regular and distinctive feedback from the Keccak state to itself (see Fig.~\ref{fig:target}), and this is captured by the FANOUT feature.

For grouping, we make use of the INPUT and OUTPUT features. These determine the sequential distance a given flip-flop has to primary inputs and outputs. It can be interpreted as the number of clock cycles it takes for a flip-flop to drive a primary output or to be reached by a primary input. Grouping flip-flops based on these features is similar to levelization. Notice from Fig.~\ref{fig:target} that if the Keccak input register is assigned to a level $L_i$, the Keccak state should be in $L_{i+1}$.


\noindent \underline{\textbf{Dealing with Imprecise Grouping.}} Locating the Keccak input is relatively harder than locating the state: when grouping flip-flops as registers, non-ideal clustering may cause Alg.~\ref{alg:repqc} to return an empty list. For example, if the 64 flip-flops of interest get grouped into two registers of 32 elements each. This is a well-understood problem in RE and has also been reported in related works \cite{re_crypto}. In order to tackle this issue, we propose two variants of the REPQC algorithm, REPQC(g) and REPQC(i), where $g$ stands for groups and $i$ stands for individuals. REPQC(g) matches the algorithm described in Alg.~\ref{alg:repqc} while REPQC(i) assumes groups with a size of 1, which in practice is the same as performing individual flip-flop comparisons. The REPQC(i) variant is more time intensive because it cannot eliminate entire groups at a time (line 20 of Alg.~\ref{alg:repqc} is therefore not considered).

\subsection{HTH Architecture}
\label{subsec:hth}

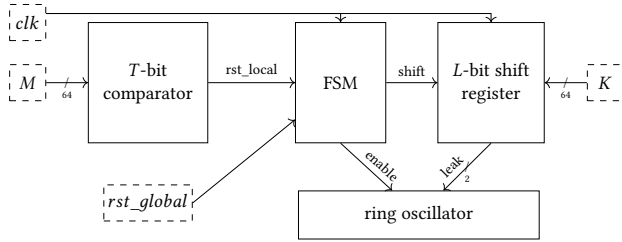
\begin{figure}
    \centering
        \begin{adjustbox}{scale=0.8}

\begin{tikzpicture}[
box/.style={rectangle,draw,inner sep=2pt,minimum width=15mm, minimum height=20mm},
largebox/.style={rectangle,draw,inner sep=2pt,minimum width=40mm, minimum height=8mm},
input/.style={rectangle,draw,dashed,inner sep=2pt,minimum size=6mm},
dot/.style={circle,inner sep=0pt,minimum size=0.5mm,draw,fill=black},
buswidth/.style={decoration={
  markings,
  mark= at position 0.5 with {\node[font=\footnotesize] {/};\node[below=1pt] {\tiny #1};}
  }, postaction={decorate}}
]

\node [input, name=clk] {$clk$};
\node [input, name=message, below of=clk] {$M$};
\node [box, name=comp, right of=message,node distance=2cm] {\begin{tabular}{c} $T$-bit \\ comparator \end{tabular}};
\node [input, name=rst, below of=comp,node distance=2cm] {$rst\_global$};
\node [box, name=fsm, right of=comp,node distance=3.2cm] {FSM};
\node [box, name=sr, right of=fsm,node distance=2.5cm] {\begin{tabular}{c} $L$-bit shift \\ register \end{tabular}};
\node [input, name=message2, right of=sr, node distance=1.9cm] {$K$};
\node [largebox, name=ro, below of=fsm,node distance=2.2cm,xshift=13mm] {ring oscillator};

\draw [->,buswidth={64}] (message.east) -- (comp);
\draw [->,buswidth={64}] (message2.west) -- (sr);
\draw [->] ([yshift=+1.5mm]clk.east) -| (fsm);
\draw [->] ([yshift=+1.5mm]clk.east) -| (sr);
\draw [->] (rst.east) -- (fsm);
\draw [->] (comp.east) -- (fsm.west) node [pos=0.50,above,font=\footnotesize] {rst\_local};
\draw [->] (fsm.east) -- (sr.west) node [pos=0.50,above,font=\footnotesize] {shift};
\draw [->] (fsm.south) -- (ro) node [pos=0.65,above,sloped,font=\footnotesize] {enable};
\draw [->,buswidth={2}] (sr.south) -- (ro) node [pos=0.65,above,sloped,font=\footnotesize] {leak};

\end{tikzpicture}
\end{adjustbox}
\vspace{-10pt}
\caption{High-level diagram of the HTH. 
}
    \label{fig:trojan_arch}
\end{figure}

First, let us describe the chosen HTH architecture, which is amenable to both high-speed and lightweight PQC accelerators. The high-level diagram of the HTH architecture is depicted in Fig.~\ref{fig:trojan_arch} and is employed by both \texttt{A-IP} and \texttt{A-FO} adversaries. The inputs $M$, $K$, $clk$, and $rst\_global$ are marked with dashed boxes. The HTH consists of a multi-bit comparator, a control logic implemented by a Finite State Machine (FSM), a shift register, and a ring oscillator. 

\begin{defi}{Working principle} The HTH awaits for a carefully selected message $M$ to be sent for hashing (line 3 in Alg.~\ref{alg:Dil_sign}). Once this happens, the HTH is triggered and the FSM is reset using $rst\_local$. Knowing that $K$ will be hashed next (line 4 in Alg.~\ref{alg:Dil_sign}), the FSM loads $K$ before it gets fully absorbed by Keccak. Next, the FSM activates the ring oscillator via its $enable$ pin. The content of the shift register is shifted whenever $shift$ is asserted. At every clock cycle, two bits of $K$ appear on $leak$. The FSM stops after the shift register is empty.
\end{defi}

For scalability, the HTH has two parameters, $T$ and $L$, where $T$ determines the size of the trigger (which determines the comparator size). The parameter $L$ determines the size of the shift register, meaning how many bits of $K$ will be stored before they are leaked. The higher the values of $T$ and $L$, the bigger the HTH. $K$ can also be leaked in parts that can be indexed using leftover bits of message $M$. So the buffer storing $K$ can be extremely small or large depending on the bits it has to leak. This flexibility enables concealed trojan insertion in not just high-speed but also extremely lightweight designs. In our repository, we provide all combinations of $T=\{16,32,64\}$ and $L=\{16,32,64\}$ for nine HTHs that leak entire secret $K$ at once. In the results that follow, we set $T=64$ without loss of generality since this is the largest and most difficult HTH to insert.

\begin{figure}
    \centering
    \begin{adjustbox}{scale=0.9}
    
    \begin{circuitikz}
     \tikzset{mux 4by2/.style={muxdemux, muxdemux def={Lh=4, NL=4, Rh=3,  NB=2, w=2, square pins=1}}}
     \ctikzset{
    	logic ports=ieee,
    	logic ports/scale=0.5,
    	logic ports/fill=white,
        fill=white,
    }
    \node[nand port] (nand) at (0,0){};
    
    \node[not port] (not1) at (1,0){};
    \node[not port] (not2) at (2,0){};
    \node[not port] (not3) at (3,0){};
    \node[not port] (not4) at (4,0){};
    
    \draw (nand.out) -- (not1.in);
    \draw (nand.in 2) -- ++(-0.0,0)node[left](enable){enable};
    \draw (not1.out) -- (not2.in);
    \draw (not2.out) -- (not3.in);
    \draw (not3.out) -- (not4.in);
    
    \node[buffer port,scale=1.0] (buf1) at (5,1.2){};
    \node[buffer port,scale=1.2] (buf2) at (5,0.4){};
    \node[buffer port,scale=1.4] (buf3) at (5,-0.4){};
    \node[buffer port,scale=1.6] (buf4) at (5,-1.2){};
    
    \draw (not4.out) -- (buf1.in);
    \draw (not4.out) -- (buf2.in);
    \draw (not4.out) -- (buf3.in);
    \draw (not4.out) -- (buf4.in);
    
    \node[mux 4by2] (mux) at (6.7,0){\begin{tabular}{c} \texttt{4:1} \\ mux 
    \end{tabular}};
    
    \draw (mux.bpin 1) -- ++(-0.0,0.0)node[below](leak1){leak[1:0]};
    
    \draw (buf1.out) -- (mux.lpin 1);
    \draw (buf2.out) -- (mux.lpin 2);
    \draw (buf3.out) -- (mux.lpin 3);
    \draw (buf4.out) -- (mux.lpin 4);
    
    \draw (mux.rpin 1) --  node[anchor=east, above=2pt ] {} ++(0,1.6) -| (nand.in 1);
    
    \end{circuitikz}
    \end{adjustbox}
    \vspace{-10pt}
    \caption{Schematic of the ring oscillator. The different-sized BUFFs modulate the frequency and power consumption.}
    \label{fig:trojan_ro}
\end{figure}
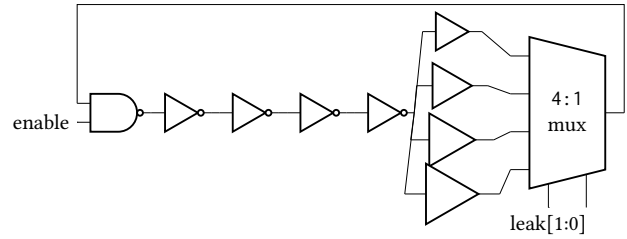

\pgfplotstableread[row sep=\\,col sep=&]{
    Circuit  & Regs  & Depends & Groups & Groups64   \\
    kali\_bl  & 12434 & 396700  & 1134 & 14 \\
    kali\_nm  & 12434 & 396700  & 1180 & 23 \\
    kali\_dc  & 11732 & 321900  & 963  & 17  \\
    kali\_ret  & 12578 & 399200  & 1175 & 21 \\
    saber &  9041 & 815200  & 421  & 21  \\
    saber\_fo & 17492 & 1084500 & 1504 & 41 \\
    gmu       & 49367 & 7956400 & 4461 & 122 \\
    }\circuits 

\pgfplotstableread[row sep=\\,col sep=&]{
    Circuit  & Regs  & Depends & Groups & Groups64   \\
    1  & 12434 & 396700  & 1134 & 14 \\
    2  & 12434 & 396700  & 1180 & 23 \\
    3  & 11732 & 321900  & 963  & 17  \\
    4  & 12578 & 399200  & 1175 & 21 \\
    5 &  9041 & 815200  & 421  & 21  \\
    6 & 17492 & 1084500 & 1504 & 41 \\
    7      & 49367 & 7956400 & 4461 & 122 \\
    }\circuitsINT

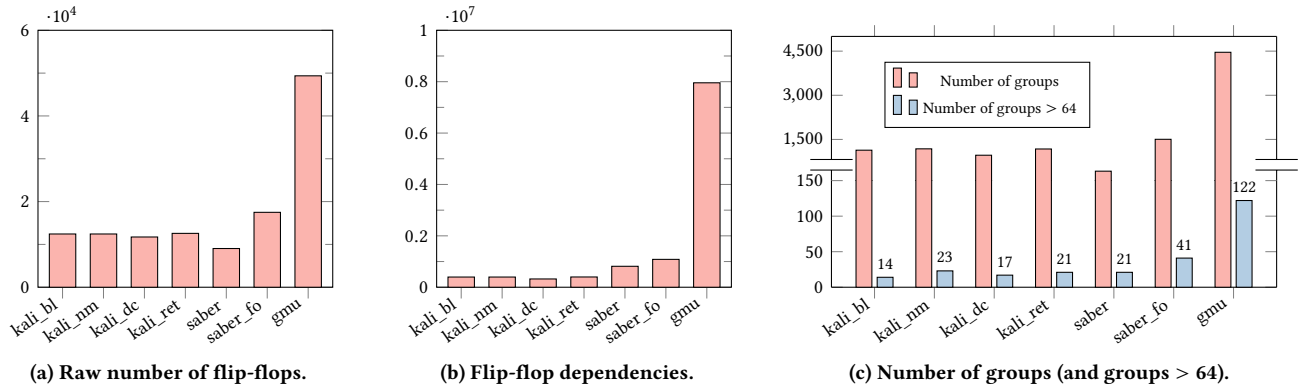
\begin{figure*}
     \centering
     \begin{subfigure}[b]{0.29\textwidth}
         \centering
            \begin{tikzpicture}
                \begin{axis}[
                        ybar,
                        symbolic x coords={kali\_bl,kali\_nm,kali\_dc,kali\_ret,saber,saber\_fo,gmu},
                        xtick=data,
                        xtick pos = bottom,
                        x tick label style={rotate=35, anchor=east,font=\footnotesize},
                        y tick label style={font=\footnotesize},                        
                        width = 5.5cm,
                        height = 5.0cm,
                        minor y tick num = 1,
                        ymin = 0,
                        ymax = 60000,                        
                    ]
                    \addplot [fill=mycolor1] table[x=Circuit,y=Regs]{\circuits};
                  \end{axis}
            \end{tikzpicture}
            \vspace{-5pt}
         \caption{Raw number of flip-flops.}
         \label{fig:raw_regs}
     \end{subfigure}
     \hfill
     \begin{subfigure}[b]{0.29\textwidth}
         \centering
         
  \begin{tikzpicture}
    \begin{axis}[
            ybar,
            symbolic x coords={kali\_bl,kali\_nm,kali\_dc,kali\_ret,saber,saber\_fo,gmu},
            xtick=data,
            xtick pos = bottom,
            width = 5.5cm,
            height = 5.0cm,
            x tick label style={rotate=35, anchor=east,font=\footnotesize},
            y tick label style={font=\footnotesize},            minor y tick num = 1,
            ymin = 0,
            ymax=10000000,
        ]
        \addplot  [fill=mycolor1]  table[x=Circuit,y=Depends]{\circuits};
      \end{axis}
\end{tikzpicture}
         \vspace{-5pt}         
         \caption{Flip-flop dependencies.}
         \label{fig:reg_deps}
     \end{subfigure}
     \hfill
     \begin{subfigure}[b]{0.4\textwidth}
         \centering

\begin{tikzpicture}
\begin{groupplot}[
    xtick={1,2,3,4,5,6,7},
    xtick pos = bottom,
    x tick label style={rotate=35, anchor=east,font=\footnotesize},
    y tick label style={font=\footnotesize},        
    ymin = 1,
    ymax = 5000, 
    xmin=1,
    xmax=7,
    ytick distance=1,
    ybar,
    every node near coord/.append style={font=\scriptsize},
    legend pos=north west,
      group style={
        group size=1 by 2,
        xticklabels at=edge bottom,
        vertical sep=0pt,
      },
      legend style={
        font=\scriptsize,
        at={(1.5cm,3.0cm)},
        anchor=north,
        legend columns=1,
      },
        /tikz/bar width=6pt,
        enlarge x limits=0.12,
    ]
    \nextgroupplot[ 
      width = 7.5cm,
      height = 3.5cm,
      ymin=100,ymax=5000,
      ytick={1500,3000,4500},
      axis x line*=top,
      axis y discontinuity=parallel,
      %
      y filter/.code={
        \pgfmathifthenelse{#1 < 123}{NaN}{#1},
      },
    ]        
        \addplot [fill=mycolor1] table[x=Circuit,y=Groups]{\circuitsINT};
        \addplot [fill=mycolor2] table[x=Circuit,y=Groups64]{\circuitsINT};
     
    \nextgroupplot[ 
      ymin=0,ymax=150,
      ytick distance=50,
       width = 7.5cm,
       height = 3cm,
      axis x line*=top,
      xticklabels={kali\_bl,kali\_nm,kali\_dc,kali\_ret,saber,saber\_fo,gmu},
        xtick pos = bottom,
      xmin=1,
      xmax=7,
      nodes near coords,
    ]        
        \addplot [fill=mycolor1] table[x=Circuit,y=Groups]{\circuitsINT};
        \addplot [fill=mycolor2] table[x=Circuit,y=Groups64]{\circuitsINT};
        
        \legend{Number of groups, Number of groups $>$ 64}
\end{groupplot}
\end{tikzpicture}    
         \vspace{-5pt}
         \caption{Number of groups (and groups $>$ 64).}
         \label{fig:groups}
     \end{subfigure}
     \vspace{-10pt}
        \caption{Characteristics of the circuits considered in this work.}
        \label{fig:re_characteristics}
\end{figure*}

 Fig.~\ref{fig:trojan_ro} illustrates details of the ring oscillator depicted in Fig.~\ref{fig:trojan_arch}. By far, the ring oscillator is the most important part of the payload of the HTH. The proposed ring oscillator is capable of tuning its own frequency: depending on the value of $leak$, different paths will be multiplexed for the feedback path of the oscillator. By selecting different paths, the frequency of oscillation becomes modulated by $leak$, and so does the power consumption of the oscillator. This is used by the adversary to leak information via a (power) side channel. More details about the operation of the oscillator are given in Appendix~\ref{app:osc}.

    \section{Results and Discussions}
\label{sec:results}

\subsection{Considered PQC Accelerators for Experiments}
In our RE efforts, we considered seven unique PQC accelerators with different characteristics and different implementors behind them. Before delving into the more specific details of the seven selected PQC accelerators, we want to clarify that there is no particular fitness criteria behind their selection -- We essentially considered every open source PQC hardware accelerator design that we were able to source. Since all of them make use of Keccak by definition, all of them can be analyzed with REPQC. More specific details of the chosen accelerators are given below.

First, we consider \emph{kali} \cite{kali}, a PQC accelerator with a crafty combination of Dilithium and Kyber. When attacking \emph{kali}, we target only its Dilithium functionality. Since the design also supports Kyber, it complicates the RE effort. We consider four variants of \emph{kali}: \emph{kali\_bl} is the original baseline design \cite{kali}; In \emph{kali\_nm}, we remove the memories from the design to emulate a harder scenario in which the adversary cannot use the memories for netlist understanding; In \emph{kali\_dc}, we perform synthesis using Synopsys Design Compiler, whereas all other circuits are synthesized in Cadence Genus; Finally, in \emph{kali\_ret}, retiming\footnote{Retiming is a circuit optimization technique that shifts the location of combinational logic with respect to sequential logic in order to balance the logic stages in a design.
} is performed. Supposedly, retiming could reshape the flip-flop dependencies and make our fanin/fanout observations invalid.

We also consider two other PQC designs that implement Saber, a KEM algorithm that also makes use of Keccak. The reasoning is that if one can find Keccak in an accelerator for a completely different algorithm, this is an indication that Keccak can, in fact, be used as an anchor for reverse engineering. Furthermore, we consider the Saber accelerator described in \emph{saber} \cite{saber_jcen}. This accelerator employs Keccak in a slightly different manner than \emph{kali}, i.e., different rates and capacities are employed. For certain, the use of different rates and capacities has no bearing on the permutation functions, so the ability of REPQC to find the Keccak state should be precisely the same. We validate this premise by including this accelerator. In \emph{saber\_fo} \cite{masked_keccak}, a first-order masked version of Keccak is considered to, potentially, further increase the difficulty of RE since this implementation contains many more flip-flops and intricate connections between them. Finally, in the \emph{gmu} design \cite{gmu}, multiple instances of Keccak are part of the same Dilithium accelerator, creating a significant challenge for RE. 

For all experiments reported in this work, logic synthesis was executed in Cadence Genus v19.10, physical synthesis using Cadence Innovus v18.10, and the technology considered is a commercial 28nm CMOS technology. Several implementation scripts are available in our repository for the sake of reproducibility. All execution times are reported in a Xeon CPU running at 3.60GHz.

\subsection{Experimental Results}
\label{subsec:results}


\noindent \underline{\textbf{Reverse Engineering with REPQC.}}
Let us describe the results related to the reverse engineering aspects of our work. In Fig.~\ref{fig:re_characteristics}, the characteristics of the seven studied circuits are presented, and Fig.~\ref{fig:raw_regs} gives the total number of flip-flops per circuit; notice that the $gmu$ circuit is significantly larger than the others. Fig.~\ref{fig:reg_deps} depicts the number of flip-flop dependencies, indicating that the $gmu$ circuit is much more complex than the others. Finally, in Fig.~\ref{fig:groups}, the number of groups is shown (melon color), as well as the number of groups with more than 64 flip-flops (steel blue color). The difference between the  bars gives a notion of the difficulty of the problems tackled by REPQC(g) and REPQC(i). Notice that the vertical axis has a discontinuity and different scales.
   
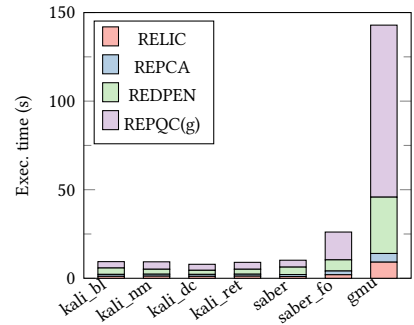
\begin{figure}
\centering
\begin{tikzpicture}
    \begin{axis}[
        ybar stacked,
        symbolic x coords={kali\_bl,kali\_nm,kali\_dc,kali\_ret,saber,saber\_fo,gmu},
        xtick=data,
        xtick pos = bottom,
        x tick label style={rotate=35, anchor=east,font=\footnotesize},
        y tick label style={font=\footnotesize},
        ylabel style={font=\footnotesize},
        width = 0.7\columnwidth,
        minor y tick num = 1,
        ymin = 0,
        ymax = 150,
        ylabel=Exec. time (s),
        legend pos=north west,
        legend style={font=\footnotesize},
        ]
        \addplot[fill=mycolor1] table[x=Circuit,y=RELIC]{\exectime};
        \addplot[fill=mycolor2] table[x=Circuit,y=RELICPCA]{\exectime};
        \addplot[fill=mycolor3] table[x=Circuit,y=REDPEN]{\exectime};
        \addplot[fill=mycolor4] table[x=Circuit,y=REPQCSTRATG]{\exectime};
        \legend{RELIC, REPCA, REDPEN, REPQC(g)}
    \end{axis}
\end{tikzpicture}
\vspace{-10pt}
\caption{Execution time of the many RE steps.}
\label{fig:exec_g}
\end{figure}

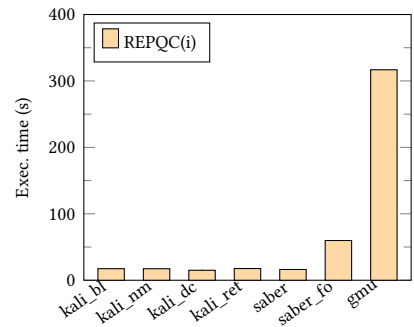
\begin{figure}
\centering
\begin{tikzpicture}
    \begin{axis}[
        ybar stacked,
        symbolic x coords={kali\_bl,kali\_nm,kali\_dc,kali\_ret,saber,saber\_fo,gmu},
        xtick=data,
        xtick pos = bottom,
        x tick label style={rotate=35, anchor=east,font=\footnotesize},
        y tick label style={font=\footnotesize},
        ylabel style={font=\footnotesize},
        width = 0.7\columnwidth,
        minor y tick num = 1,
        ymin = 0,
        ymax = 400,
        ylabel=Exec. time (s),
        legend pos=north west,
        legend style={font=\footnotesize},
        ]
        \addplot[fill=mycolor5] table[x=Circuit,y=REPQCSTRATI]{\exectime};
        \legend{REPQC(i)}
    \end{axis}
\end{tikzpicture}
\vspace{-10pt}
\caption{Execution times for REPQC(i), assuming the REPQC(g) strategy is not successful.}
\label{fig:exec_i}
\end{figure}

Execution times of the different reverse engineering steps are given in Fig.~\ref{fig:exec_g}. Notice how the execution time is dominated by REPQC(g), and the largest circuit takes approximately two minutes to be analyzed. When grouping is not reliable, the REPQC(i) strategy is utilized, and the execution times increase accordingly, as shown in Fig.~\ref{fig:exec_i}.

\pgfplotstableread[row sep=\\,col sep=&]{
    Circuit   & RELIC & RELICPCA & REDPEN & REPQCSTRATG & REPQCSTRATI   \\
    kali\_bl  & 1.1   & 1.1      & 3.6    & 3.6         & 17.4 \\
    kali\_nm  & 1.2   & 1.1      & 2.8    & 4.2         & 17.3 \\
    kali\_dc  & 1.1   & 1.1      & 2.3    & 3.4         & 15.0 \\
    kali\_ret  & 1.2   & 1.1      & 2.8    & 3.9         & 17.6 \\
    saber & 1.0   & 1.1      & 4.2    & 3.9         & 16.2 \\
    saber\_fo & 2.0   & 2.1      & 6.3    & 15.7        & 59.8 \\
    gmu       & 9.1   & 4.9      & 31.8   & 97.1        & 316.8 \\
    }\exectime 

The outcome of the REPQC analysis is provided in Tab.~\ref{tab:ckff}, where we indicate how many flip-flops are identified as Keccak state candidates. Initially, a naive search is conducted where $FIF=33$; $FOF=33$; $FIC$ and $FOC$ are unbounded (see Fig.~\ref{fig:keccak_state_filled} for details). This approach filters out thousands of flip-flops but always returns more candidates than the expected 1600 (or multiples of). In other words, the naive search finds all the expected correct candidates and a few more. A clever search procedure can be executed by setting $FOC=FIC$ and repeatedly incrementing $FOC$ until the number of candidates exceeds the expected 1600. Now, the number of candidates is much closer to the expected value. 

\begin{table*}
\small
\caption{Results of the REPQC analysis.}\vspace{-10pt}
\label{tab:ckff}
  \centering
  \begin{tabular}{lc|ccc|ccc}
  \hline
  \multicolumn{2}{c|}{} & \multicolumn{3}{c|}{\textbf{Naive search}} & \multicolumn{3}{c}{ \textbf{Clever search}} \\ 
\thead{Circuit} & \thead{Flip-flops} & \thead{FIF, FIC} & \thead{FOF, FOC} & \thead{CKFF/KFF} & \thead{FIF, FIC} & \thead{FOF, FOC} & \thead{CKFF/KFF} \\ \hline
$kali\_bl$  & 12435 & $[33,\infty]$ & $[34,\infty]$ & 1774/1600 & $[34,\infty]$ & $[34,35]$ & 1603/1600 \\ 
$kali\_nm$  & 12435 & $[33,\infty]$ & $[34,\infty]$ & 1774/1600 & $[34,\infty]$ & $[34,35]$ & 1603/1600 \\ 
$kali\_dc$  & 11733 & $[33,\infty]$ & $[34,\infty]$ & 1709/1600 & $[34,\infty]$ & $[34,35]$ & 1603/1600 \\ 
$kali\_ret$  & 12579 & $[33,\infty]$ & $[34,\infty]$ & 1772/1600 & $[34,\infty]$ & $[34,35]$ & 1603/1600 \\  
$saber$ &  9042 & $[33,\infty]$ & $[34,\infty]$ & 1868/1600 & $[34,\infty]$ & $[34,35]$ & 1720/1600 \\  
$saber\_fo$ & 17493 & $[33,\infty]$ & $[34,\infty]$ & 3205/3200 & $[34,\infty]$ & $[34,145]$ & \textbf{3200/3200} \\  
$gmu$       & 49368 & $[33,\infty]$ & $[34,\infty]$ & 5794/4800 & $[34,\infty]$ & $[34,35]$ & 4861/4800 \\ \hline 
\multicolumn{8}{l}{\footnotesize{FIF and FIC stand for FanIn Floor and FanIn Ceiling. Similarly, FOF and FOC stand for FanOut Floor and FanOut Ceiling. }} \\
\multicolumn{8}{l}{\footnotesize{KFF stands for Keccak state flip-flops. CKFF stands for candidate Keccak state flip-flops. CKFF includes all the KFF candidates.}}
\end{tabular}

\end{table*}

For the $saber\_fo$ circuit, the Keccak state identification is perfect. For the other circuits, even if the state identification is not perfect, this is already sufficient for REPQC to succeed because \textbf{all} of the correct candidates remain in the set of candidates. A simple approach for further filtering out candidates would be to find the dominant Z-score among the candidates and eliminate those with a higher Z-score.

\noindent \underline{\textbf{HTH Insertion after REPQC.}}
For HTH insertion, we will consider two implementations of the $kali$ design, area-efficient and speed efficient, with target frequencies of 2.00GHz and 2.19GHz, respectively. First, we implement the layout of these two baseline designs by performing placement, clock tree synthesis, optimizations, and routing. The results are summarized in Tab.~\ref{tab:comparison_small} and Tab.~\ref{tab:comparison_big} for the area and speed-efficient implementations, respectively. Implementation scripts and floorplan decisions are provided in our repository. Both baseline designs have been generated using a commercial 28nm CMOS technology and an industry-grade design flow; therefore, they are considered fabrication-ready since they pass all design rule checks. We also highlight that the clock frequency of these designs is a near match to the one reported by the authors of \cite{kali}.

\begin{table*}
\small

\caption{Characteristics of the $kali$ design when implemented with a very tight area.}\vspace{-10pt}
\label{tab:comparison_small}
\centering
\begin{tabular}{l|c|c|c|c|c|c|c}
\hline
\thead{Design } & \thead{Trojan} & \thead{Dimensions \\($\mu m$ x $\mu m$)} & \thead{Density\\ (\%)} & \thead{Cell\\ count} & \thead{Frequency\\ ($GHz$)} & \thead{Static power\\ ($\mu W$)} & \thead{Power\\ ($\mu W$)} \\ \hline 
Baseline ($kali\_bl$)                 & N/A        & 460x670 & 66.70\% & 58215 & 2.00 & 3.860 & 354.85 \\ 
\texttt{A-IP} compromised & $T$=64, $L$=16 & 460x670 & 67.83\% & 59026 & 2.00 & 3.929 & 360.37 \\ 
\texttt{A-IP} compromised & $T$=64, $L$=32 & 460x670 & 68.79\% & 59218 & 2.00 & 3.982 & 363.06 \\ 
\texttt{A-IP} compromised & $T$=64, $L$=64 & 460x670 & 69.58\% & 58747 & \textbf{1.96} & 3.985 & 364.77 \\ 
\texttt{A-FO} compromised & $T$=64, $L$=16 & 460x670 & 66.87\% (+0.17\%) & 58383  & 2.00 & 3.868 & 355.11 \\ 
\texttt{A-FO} compromised & $T$=64, $L$=32 & 460x670 & 66.92\% (+0.22\%) & 58415  & 2.00  & 3.871 & 355.19 \\ 
\texttt{A-FO} compromised & $T$=64, $L$=64 & 460x670 & 67.01\% (+0.31\%) & 58485  & 2.00  & 3.877 & 355.37 \\\hline 
\end{tabular}

\end{table*}

\begin{table*}[t]
\small

\caption{Characteristics of the $kali$ design when implemented with relaxed area and challenging frequency.}\vspace{-10pt}
\label{tab:comparison_big}
\centering
\begin{tabular}{l|c|c|c|c|c|c|c}
\hline
\thead{Design} & \thead{Trojan} & \thead{Dimensions \\($\mu m$ x $\mu m$)} & \thead{Density\\ (\%)} & \thead{Cell\\ count} & \thead{Frequency\\ ($GHz$)} & \thead{Static power\\ ($\mu W$)} & \thead{Power\\ ($\mu W$)} \\ \hline 
Baseline ($kali\_bl$)                 & N/A        & 540x670 & 43.70\% & 59880 & 2.19 & 4.084 & 392.57 \\  
\texttt{A-IP} compromised & $T$=64, $L$=16 & 540x670 & 43.59\% & 59413 & \textbf{2.17} & 4.042 & 387.18\\ 
\texttt{A-IP} compromised & $T$=64, $L$=32 & 540x670 & 43.83\% & 59952 & \textbf{2.11} & 4.073 & 385.76\\ 
\texttt{A-IP} compromised & $T$=64, $L$=64 & 540x670 & 44.23\% & 60367 & \textbf{2.18} & 4.138 & 391.11\\ 
\texttt{A-FO} compromised & $T$=64, $L$=16 & 540x670 & 43.80\% (+0.10\%) & 60048 & 2.19 & 4.092 & 392.88 \\ 
\texttt{A-FO} compromised & $T$=64, $L$=32 & 540x670 & 43.83\% (+0.13\%) & 60080 & 2.19 & 4.095 & 393.00 \\ 
\texttt{A-FO} compromised & $T$=64, $L$=64 & 540x670 & 43.89\% (+0.19\%) & 60150 & 2.19 & 4.101 & 393.27 \\\hline 

\end{tabular}
\end{table*}

A few entries on these tables are marked in bold to indicate scenarios in which the presence of an HTH has made the design less efficient than the baseline, which is not desirable from the point of view of an adversary. We note that this is the case for all \texttt{A-IP} designs in Tab.~\ref{tab:comparison_small}, highlighting yet another disadvantage of \texttt{A-IP} over \texttt{A-FO}. For this reason, we envision that the adversary might have to iterate a few times until he/she is satisfied with the `quality' of the HTH. In Fig.~\ref{fig:flowchart}, this is visualized with the diamond-shaped decision block ``Fits?''.

The reason why the HTHs inserted by \texttt{A-FO} are less intrusive to the design is the ECO flow that the adversary uses. We clarify that the ECO flow was conceived for hotfixing a layout, such that a last-minute bug found in a design can be fixed without designers having to go through the entire flow. The ECO flow does that by promoting only local changes in a layout and attempts to keep the unaffected logic/wires untouched. This is why all vendors offering modern chip design tools support this benign functionality. Here, we turn this functionality into a malicious asset for the adversary.

Some clear trends can be observed in Tab.~\ref{tab:comparison_small}, where all \texttt{A-IP} and \texttt{A-FO} compromised designs lead to an increase in density, cell count, static power, and total power. This is expected. For the increase in power, we clarify that the numbers are reported assuming the oscillator is not enabled, which is how the victim would perceive the circuit/chip. Therefore, the increase in power is due to the HTH comparator that is constantly looking for the message $M$.
In Tab.~\ref{tab:comparison_big}, the trends are not so clear because the \texttt{A-IP} compromised designs do not meet the target frequency of 2.19GHz. The most remarkable results are those from \texttt{A-FO} with $T=64$ and $L=64$, which display a modest increment in density and power while achieving the malicious goals of the adversary. The execution time for the ECO insertion of this HTH only takes 73s.


\begin{figure}
     \centering
     \begin{subfigure}[b]{0.23\textwidth}
         \centering

\begin{tikzpicture}
    \node (n0) { \includegraphics[height=137pt]{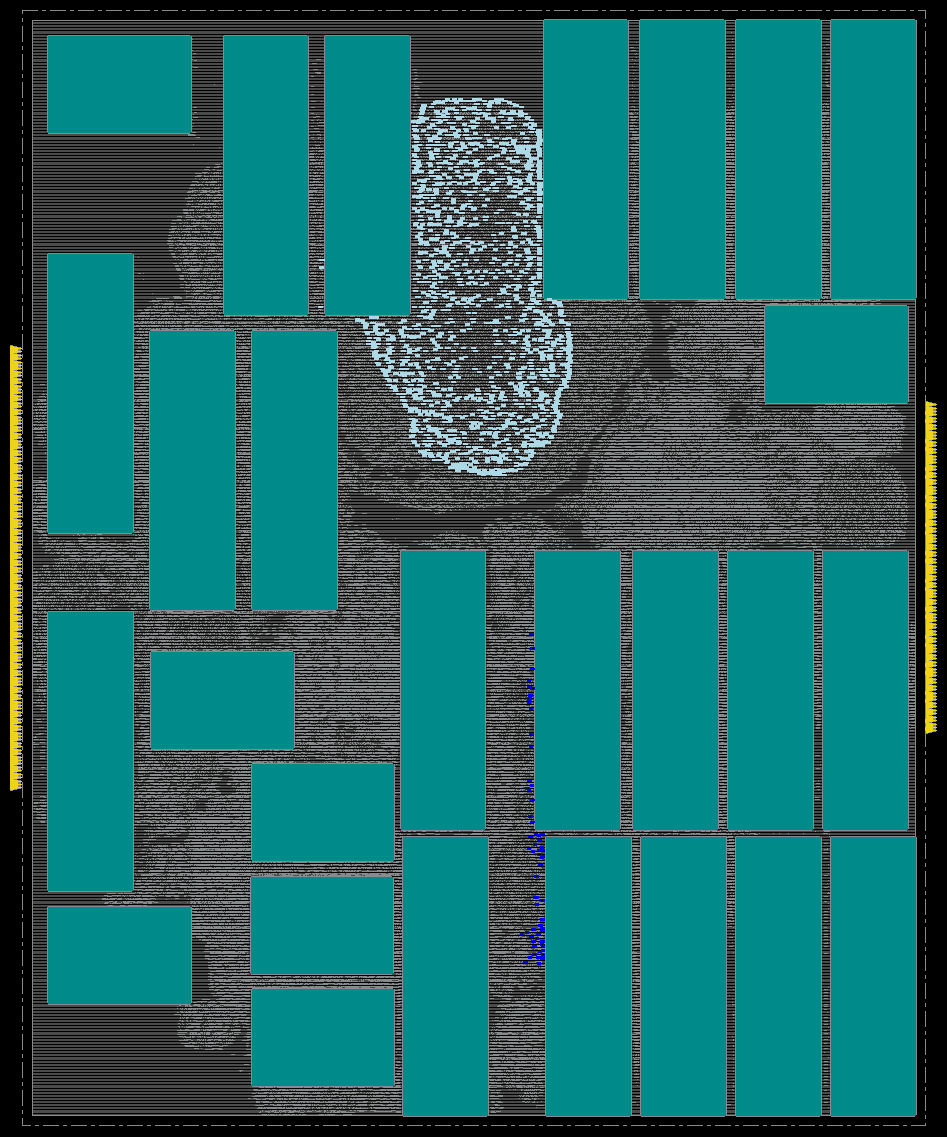} };
    \node (smallbox) at ([shift={(0.15,-1.48)}]n0.center) [black,draw,thick,minimum width=0.5cm,minimum height=0.5cm] {};
    \node (bigbox) at (3.5, 1.0) [black,draw,thick,minimum width=2cm,minimum height=2cm,inner sep=0.5pt] {\includegraphics[trim=320 130 290 620,clip,width=2cm]{figs/kali_highlighted}};    
    \draw[densely dashed, black]   (smallbox.north west) -- (bigbox.north west)
                        (smallbox.south east) -- (bigbox.south east);
\end{tikzpicture}

            \caption{Floorplan.}
            \label{fig:kali_fplan}
     \end{subfigure}
     
     \begin{subfigure}[b]{0.23\textwidth}
         \centering
          \includegraphics[height=140pt]{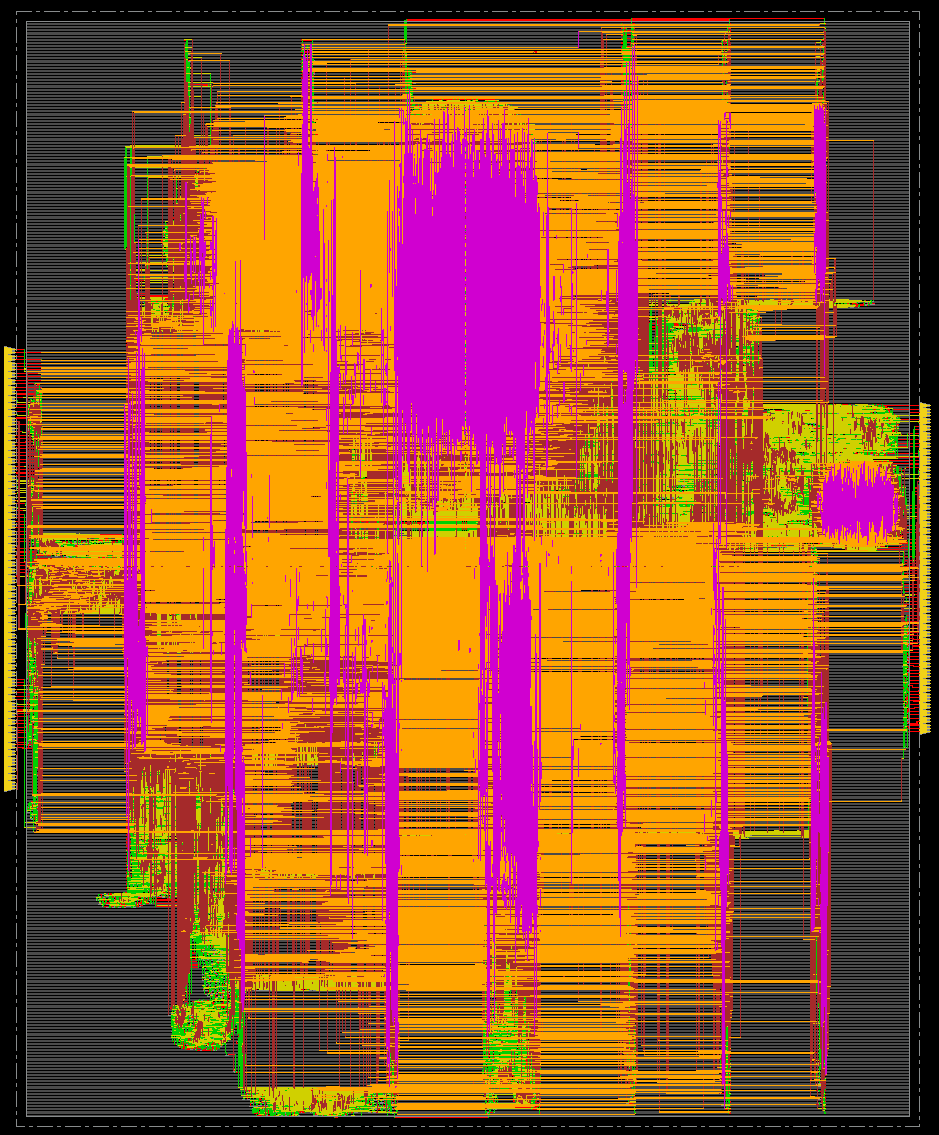}
         \caption{Routed view.}
         \label{fig:kali_routed}
     \end{subfigure}
     \hfill
     \begin{subfigure}[b]{0.23\textwidth}
         \centering
         \includegraphics[height=140pt]{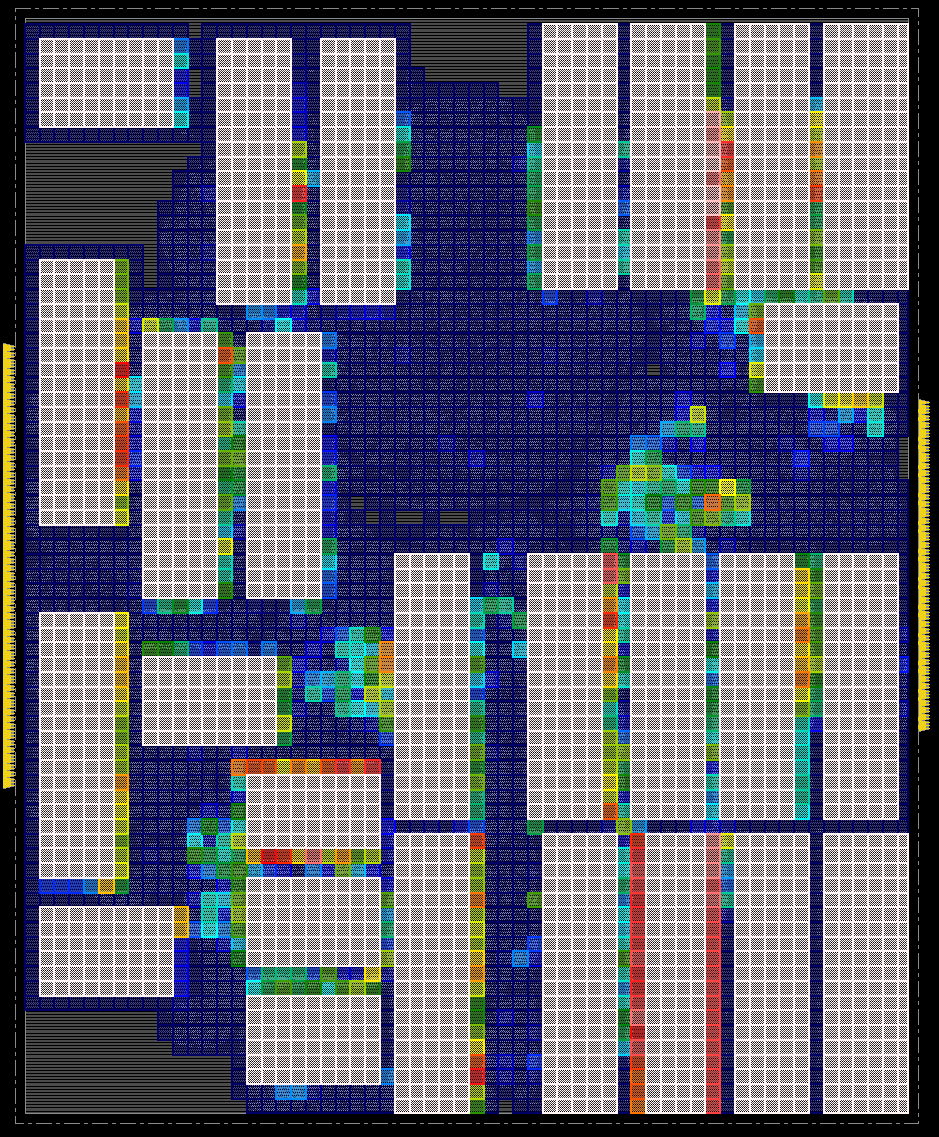}
        \caption{Placement density.}
        \label{fig:kali_density}
     \end{subfigure}
     \vspace{-10pt}
        \caption{Multiple views of the speed-efficient $kali$ design that measures 540$\mu m$ x 670$\mu m$. The Keccak state is marked in light blue, and the inputs to Keccak are marked in dark blue. Large cyan rectangles are the memories.}
        \label{fig:kali_innovus}
\end{figure}

The final layout of the speed-efficient $kali$ design is given in Fig.~\ref{fig:kali_innovus}. Fig.~\ref{fig:kali_fplan} shows the floorplan of the design where the memories are the large cyan rectangles, the Keccak state flip-flops are highlighted in light blue (paramecium-like structure), and the inputs to Keccak are marked in dark blue. Notice that the design is memory-heavy, and yet memories have no bearing on the success of REPQC. The signal routing for this design is shown in Fig.~\ref{fig:kali_routed}. Note that the area next to the Keccak input flip-flops is very congested -- pink is the highest metal in the metal stack, and it is heavily utilized in the considered region. Finally, in Fig.~\ref{fig:kali_density}, we highlight the placement density of the design, which also indicates that the region near the Keccak input flip-flops is heavily occupied. 

\begin{figure}
     \centering
     \begin{subfigure}[b]{0.23\textwidth}
         \centering
           \includegraphics[height=140pt]{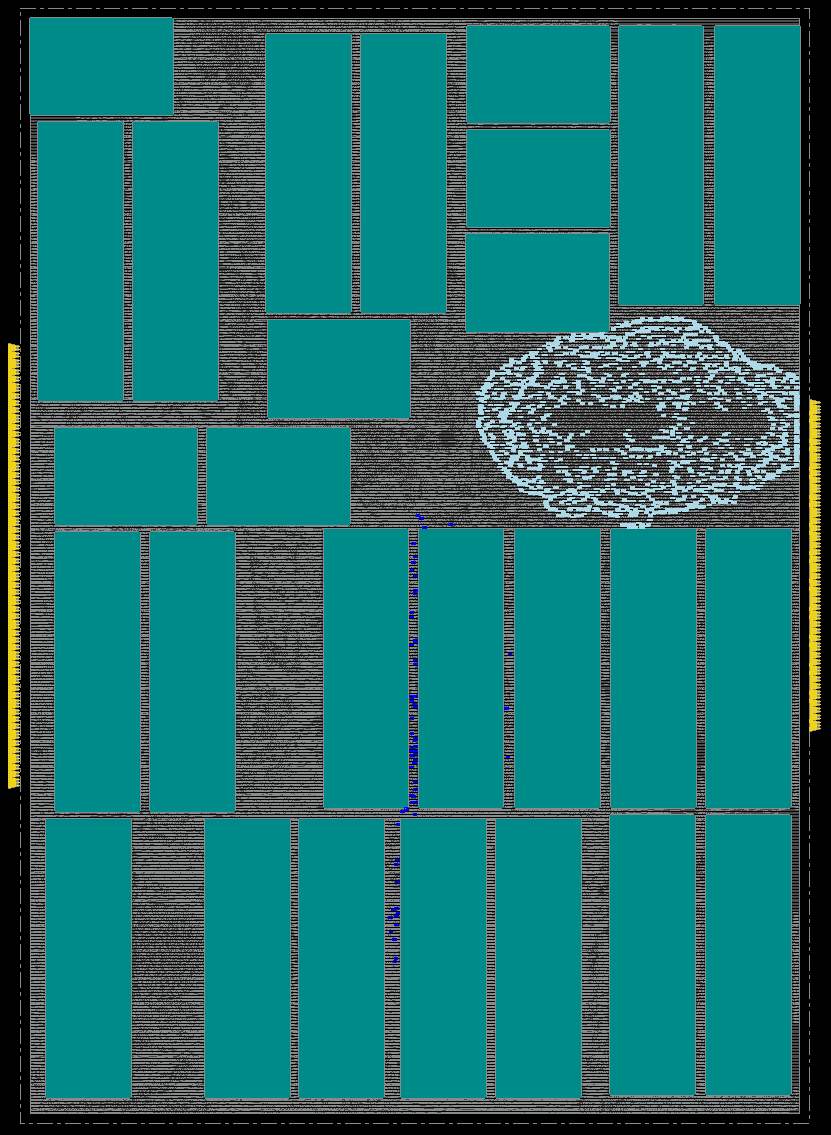}
            \caption{Floorplan w/o HTH.}
            \label{fig:kali_small_fplan}
     \end{subfigure}
     \begin{subfigure}[b]{0.23\textwidth}
         \centering
          \includegraphics[height=140pt]{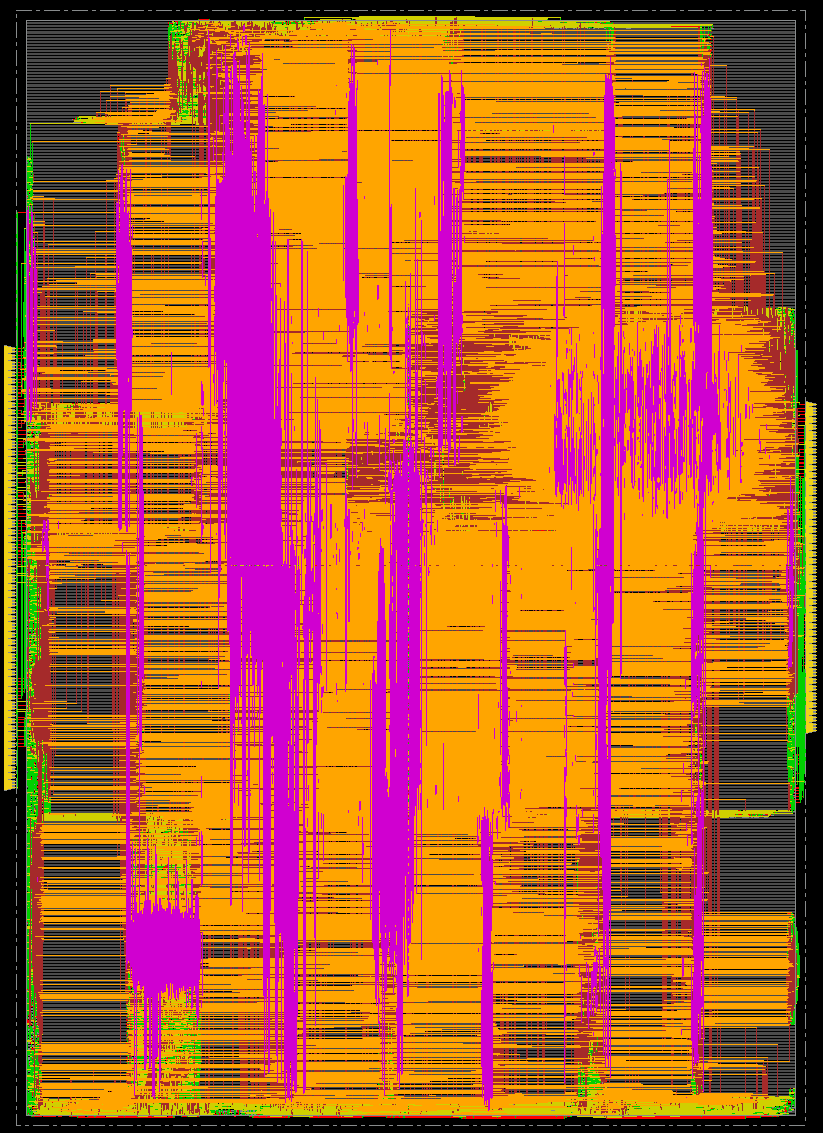}
         \caption{Routed view w/o HTH.}
         \label{fig:kali_small_routed}
     \end{subfigure}
     \begin{subfigure}[b]{0.23\textwidth}
         \centering
           \includegraphics[height=140pt]{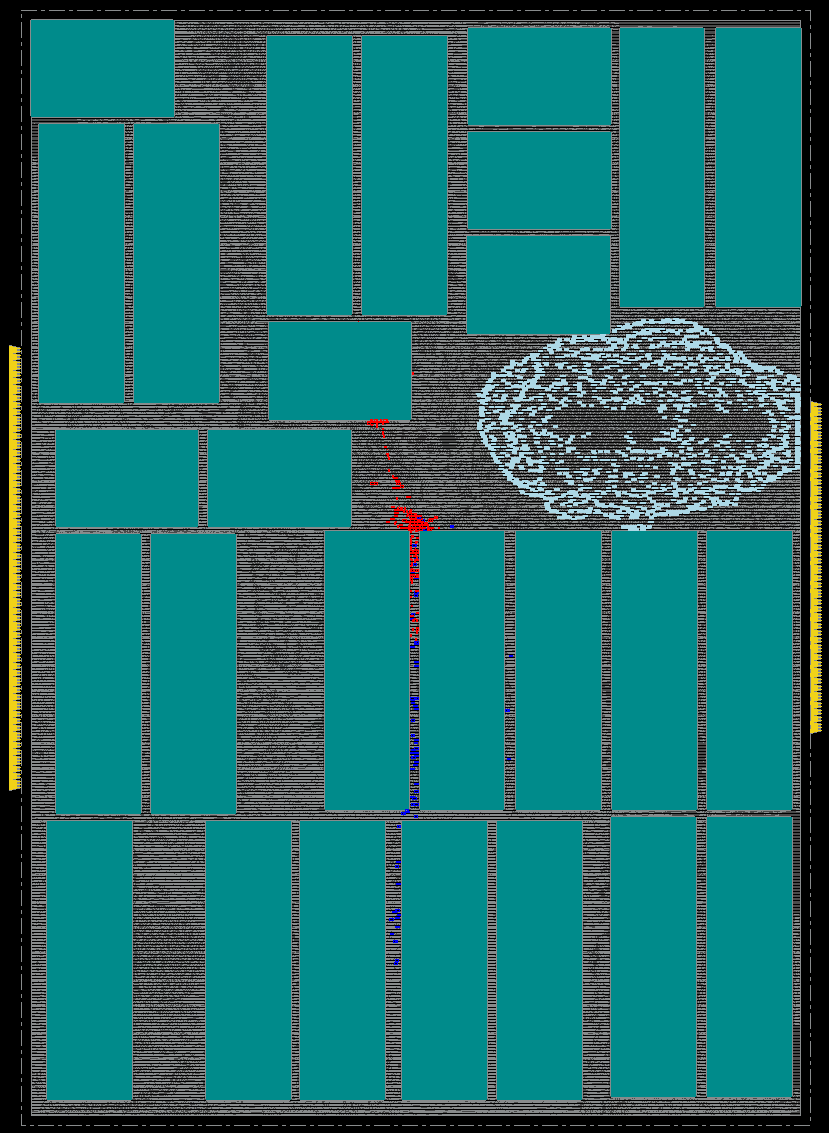}
            \caption{Floorplan with HTH.}
            \label{fig:kali_small_trojan_fplan}
     \end{subfigure}
     \begin{subfigure}[b]{0.23\textwidth}
         \centering
          \includegraphics[height=140pt]{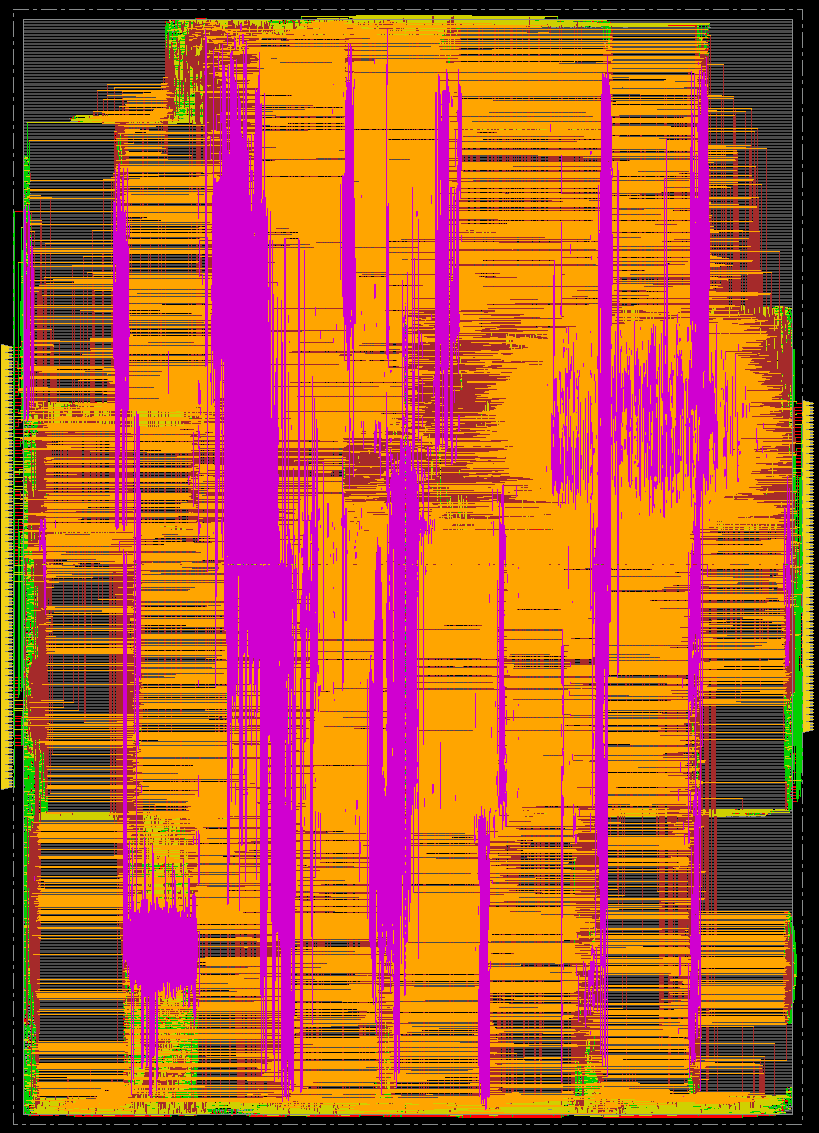}
         \caption{Routed view with HTH.}
         \label{fig:kali_small_trojan_routed}
     \end{subfigure}
     %
        \vspace{-10pt}
        \caption{Multiple views of the area-efficient $kali$ design that measures 460$\mu m$ x 670$\mu m$. The HTH is highlighted in red.}
        \label{fig:kali_small_innovus}
\end{figure}

The nature of the HTH proposed in this work, connected to the Keccak input, dictates that the HTH would compete with the existing logic for scarce routing and placement resources. This difficulty explains, partially, why some \texttt{A-IP} compromised designs have not met the targeted frequency of operation. This problem is even more pronounced for the area-efficient version of $kali$, depicted in Fig.~\ref{fig:kali_small_innovus}. By comparing Fig.~\ref{fig:kali_small_fplan} with Fig.~\ref{fig:kali_small_trojan_fplan}, it is possible to infer that the tool tried to squeeze the HTH between two memories but, due to lack of space, ended placing the HTH nearby. By comparing Fig.~\ref{fig:kali_small_routed} with Fig.~\ref{fig:kali_small_trojan_routed}, it is barely possible to notice  changes before and after HTH insertion. This is the intended outcome of a malicious ECO operation since only deep physical inspection would reveal the HTH presence. More details about the differences in routing are provided in Appendix~\ref{app:metal}.

\subsection{Discussions}
\label{sec:discussion}


The success of the attack is intimately linked to the adversary's capability to locate the Keccak state $a$, which we hypothesize that will always be there as flip-flops. Implementing the state in SRAM memory would be incredibly inefficient since Keccak rounds have concurrent read/write from/to all of its state elements. There is also the possibility of not implementing the state and instead encoding the entire 24 rounds in a purely unrolled combinational implementation. This would immediately turn Keccak into the performance bottleneck of the entire PQC accelerator and is simply not desirable. These assertions are supported by the findings of \cite{saber_tcasII}, where the authors find that only two rounds can be unrolled before the Keccak operation slows down the entire accelerator.



\noindent \underline{\textbf{\textcolor{red}{Improving the Attack}}.}
Many additional observations can be made by an attacker attempting to improve his/her reverse engineering-based attack. In REPQC, only flip-flops are considered, but combinational logic can also be used for structural analysis. We also do not use any physical hints: an adversary can correctly assume that all 1600 Keccak state flip-flops are supposed to be clumped together, and any flip-flop outside of a certain range is not a reasonable candidate. This can be visualized in Fig.~\ref{fig:kali_fplan}.

Furthermore, an attacker can specialize REPQC for a given targeted implementation by asserting values for the fanin/fanout margins (i.e., $FIF$, $FIC$, $FOF$, and $FOC$). An attacker can also specialize the RELIC recipes for better filtering Keccak state candidate flip-flops. Our analysis and experimental results show that our proposed recipes are good enough for the generic case, but other recipes may yield better results for individual circuits.


\noindent \underline{\textbf{\textcolor{blue}{Improving the Defence}}.}
A summary of the many possible defence approaches is given in Tab.~\ref{tab:summary}. Our observation is that, among the approaches considered in this work, having multiple Keccak instances is fairly effective, as is the case in the $gmu$ design. Those do not directly impede the localization of the Keccak state but make register grouping far more difficult. In fact, for $gmu$, the REPQC(g) strategy is not suitable. First-order masking has been considered in the $saber\_fo$ design. Since the input message ($M$) is only used for hashing, it will be boolean-masked (using XOR gate). The control flow will be fixed so the attacker can insert the trojan for XORing the two shares of $M$, which will be loaded in the Keccak states for hashing. Hence, the approach is very easily applicable for masked implementation. Surprisingly, the masking structure is so regular that it made REPQC  more reliable (this is what we refer as `hints' in Tab.~\ref{tab:summary}). On the other hand, it remains to be studied what effect other masking schemes would have. At a bare minimum, we expect execution times to increase as more complicated or higher order masking schemes are considered.

\begin{table}
\small

\caption{Summary of potential mitigations.}\vspace{-10pt}
\label{tab:summary}
\centering
\begin{tabular}{l|c|c}
\hline
\multirow{2}{*}{\textbf{Approach}} & \multirow{1}{*}{\textbf{Used}} & \multirow{2}{*}{\textbf{Implication for RE}} \\  

{} & {\textbf{in TW}} & {} \\ \hline 
    
\multirow{2}{*}{Multiple instances} & \multirow{2}{*}{\ding{51}} & Increases execution time \Smiley   \\
                                    &                         & Prevents register grouping \Smiley \\ \hline
\multirow{2}{*}{Masking (FO)} & \multirow{2}{*}{\ding{51}} & Increases execution time \Smiley   \\
                                    &                         & Creates additional hints \Sadey \\ \hline
Masking (Other) & \ding{53} & Increases execution time \Smiley \\ \hline     

Diversified synth. & \ding{51} & None \Neutrey   \\ \hline 
Retiming & \ding{51} & None \Neutrey   \\ \hline 

Decoys & \ding{53} & Unknown\\ \hline
\multicolumn{3}{l}{\footnotesize{TW=this work.  Smile faces are from the point of view of the defender.}}                                
\end{tabular}

\end{table}

We have also considered circuits synthesized in tools from different vendors and concluded that the Keccak round structure is not affected, as expected, since synthesis must respect the hardware description. We have also considered retiming but to no avail. In theory, the Keccak round structure is not retiming-friendly since it contains a feedback loop. Moving the combinational logic, in this case, achieves no rebalancing of the logic stages. This can be explained by the fact that the Keccak sponge function is analogous to a pipeline with a depth of one. It is possible, however, that retiming might affect the Keccak input, but we have not witnessed it in our experiments with $kali\_ret$.   

Finally, we propose a couple of other promising approaches. We hypothesize that decoy flip-flops could break the rigid Keccak structure, but this is only of interest as long as other security properties remain in place. It is also possible to consider a combination of defences to confuse the RE process.



\section{Conclusion}\label{sec:conclusion}

Today, the transition to PQC is going full steam ahead. Partially motivated by `Store Now, Decrypt Later' attacks, we are already observing the transition to quantum-resistant cryptography before quantum computers become widely available. Being so, the need to have PQC hardware accelerators is urgent. We have shown in this paper that such accelerators can be reverse-engineered and backdoored with remarkable levels of sophistication and automation. Furthermore, since this study is the first to reveal an automated approach for RE of PQC hardware accelerators, many avenues for future research remain open, both adversarial and defensive. On the adversarial side, we confidently state that locating the Keccak state in a blind netlist can be done with high confidence. However, the same cannot be said about locating the Keccak input. Our insight is that defensive approaches that fuzzy the location of the Keccak input, directly or indirectly, will pose significant challenges to an adversary. 

 \section*{Acknowledgments}

This work was partially supported by the EC through the European Social Fund in the context of the project ``ICT programme''. It was also partially supported by European Union's Horizon 2020 research and innovation programme under grant agreement No 952252 (SAFEST). 

\section*{Data Availability}
The C++ source code and a binary for REPQC is available from our repository~\cite{OURREPO}. It also contains several shell scripts for automation, many HTH samples and netlists.

\vspace{300pt}
\pagebreak
\bibliographystyle{ACM-Reference-Format}
\bibliography{sample-base}

\appendix

\section{RELIC}\label{app:relic}

\begin{figure}
     \centering
     \begin{subfigure}[b]{0.23\textwidth}
         \centering
           \includegraphics[height=95pt]{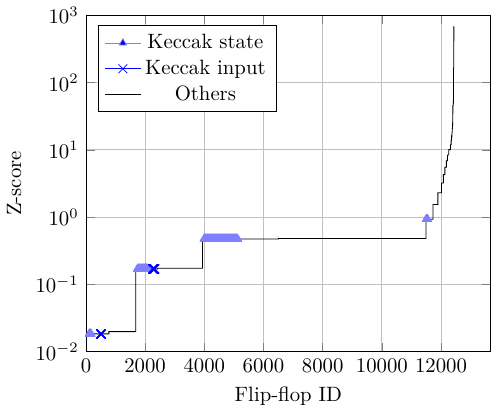}
            \vspace{-5pt}
            \caption{\emph{kali\_bl}.}
            \label{fig:zscores_panel_a}
     \end{subfigure} 
     \begin{subfigure}{0.23\textwidth}
         \centering
           \includegraphics[height=95pt]{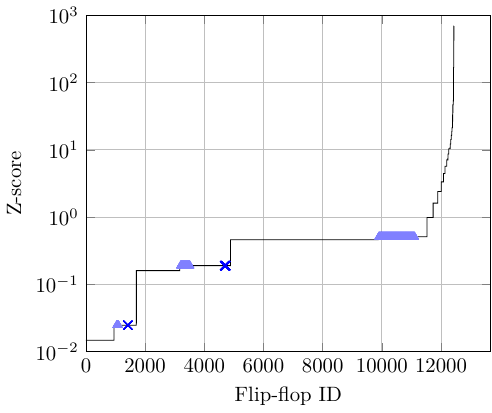}\vspace{-5pt}
         \caption{\emph{kali\_nm}.}
            \label{fig:zscores_panel_b}
     \end{subfigure}
     
     \begin{subfigure}{0.23\textwidth}
         \centering
           \includegraphics[height=95pt]{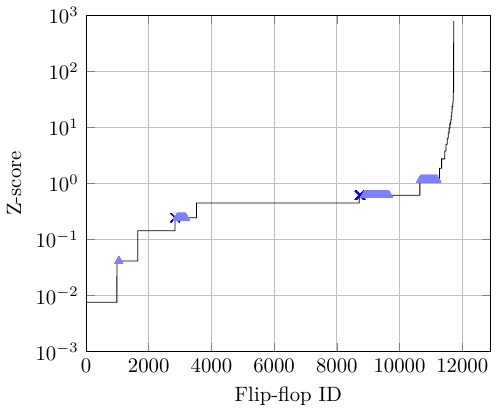}\vspace{-5pt}
        \caption{\emph{kali\_dc}.}     
            \label{fig:zscores_panel_c}
     \end{subfigure}
     \begin{subfigure}{0.23\textwidth}
         \centering
           \includegraphics[height=95pt]{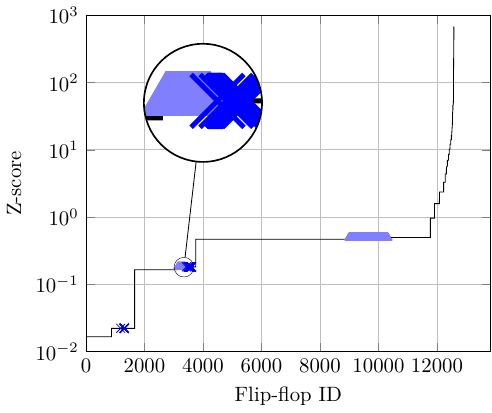}\vspace{-5pt}
            \caption{\emph{kali\_ret}.}
            \label{fig:zscores_panel_d}
     \end{subfigure}
     
     \begin{subfigure}{0.23\textwidth}
         \centering
           \includegraphics[height=95pt]{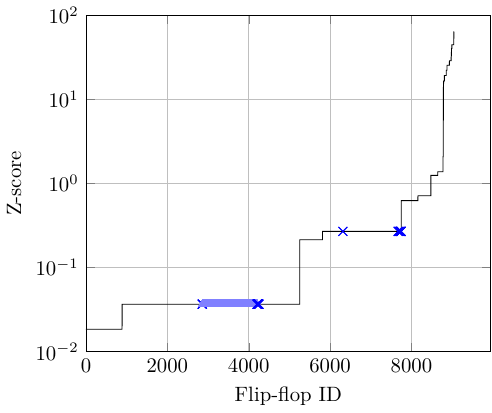}\vspace{-5pt}
         \caption{\emph{saber}.}
            \label{fig:zscores_panel_e}
     \end{subfigure}
     \begin{subfigure}{0.23\textwidth}
         \centering
           \includegraphics[height=95pt]{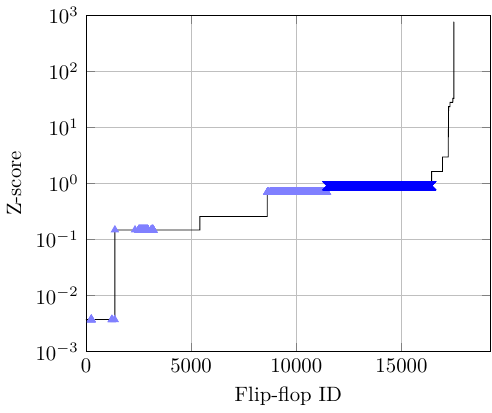}\vspace{-5pt}
        \caption{\emph{saber\_fo}.}     
            \label{fig:zscores_panel_f}
     \end{subfigure}
     
     \begin{subfigure}{0.46\textwidth}
         \centering
           \includegraphics[height=65pt]{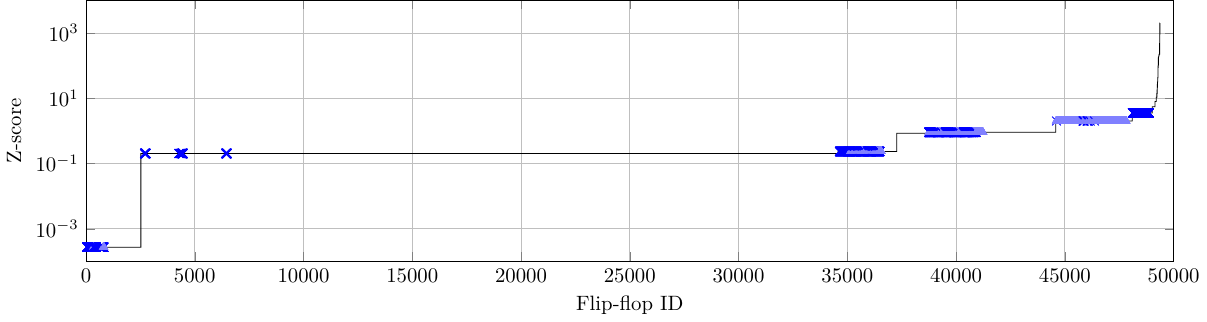}\vspace{-5pt}
        \caption{\emph{gmu}.}     
            \label{fig:zscores_panel_g}
     \end{subfigure} 
     \vspace{-10pt}
     \caption{Z-scores of the PQC accelerators considered in this work.}
     \label{fig:zscores}
\end{figure}

Each panel of Fig.~\ref{fig:zscores} is a representation of RELIC's output for a given circuit. Values close to zero indicate that the considered flip-flop is more likely to be part of a datapath than part of the control logic. The opposite is true for high values.

Notice from the distributions for $kali$ variants, i.e., panels (a-d), that there are similarities, despite these circuits have been generated by different tools and in different ways. This is an indication that the properties of Keccak are kept; thus, REPQC, which builds on RELIC, is agnostic with respect to synthesis. 

Notice too that the Keccak state flip-flops, marked in light blue triangles, do not appear continuously one after another in these plots. In particular, the plot in panel (d) reveals that the Keccak state appears in two distinct regions. However, the plot in panel (f) reveals that the state appears in 9 separate continuous regions. In other words, the Z-score alone is not sufficient to determine which flip-flops implement/do not implement the Keccak state.

\section{Oscillator}\label{app:osc}

Notice that all branches of the oscillator depicted in Fig.~\ref{fig:trojan_ro} have a common NAND and four common INVs. However, the BUFFs connected to the MUX input ports have different sizes which in turn will yield different frequencies of oscillation. Note that the NAND port controls the operation of the circuit through the external enable signal; the number of INVs determines the base oscillation frequency; the BUFFs are purposefully sized differently to create distinct power profiles from different frequencies of oscillation. 

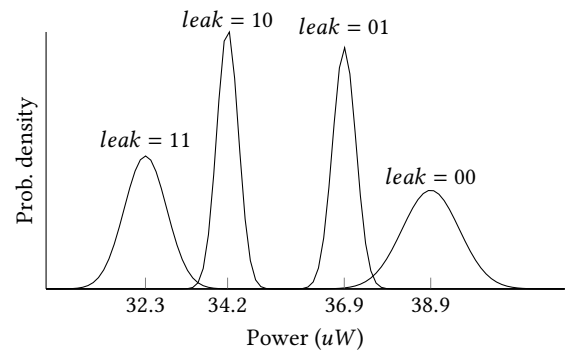
\begin{figure}
\centering
\begin{tikzpicture}[scale=1]
\begin{axis}[
  no markers, domain=30:42, samples=100,
  axis lines*=left,
  width=\columnwidth,
  height=5cm,
  xtick={32.3,34.2,36.9,38.9}, ytick=\empty,
  xlabel=Power ($uW$),
  ylabel=Prob. density,
  enlargelimits=false, clip=false, axis on top,
  ]
  \addplot [thin] {gauss(38.9, 0.66)}; 
  \addplot [thin] {gauss(36.9, 0.27)}; 
  \addplot [thin] {gauss(34.2,0.25)};       
  \addplot [thin] {gauss(32.3,0.49)};     
 
  \coordinate (A) at (axis cs:38.9,0);
  \coordinate (B) at (axis cs:36.9,0);
  \coordinate (C) at (axis cs:34.2,0);
  \coordinate (D) at (axis cs:32.3,0);


\end{axis}
\node at ([yshift=1.5cm]A) {$leak=00$};
\node at ([yshift=3.5cm]B) {$leak=01$};
\node at ([yshift=3.6cm]C) {$leak=10$};
\node at ([yshift=2cm]D) {$leak=11$};

\end{tikzpicture}
\vspace{-10pt}
\caption{Monte Carlo analysis of the ring oscillator.}
\label{fig:mc}
\end{figure}

The range of operation of the oscillator is given in Fig.~\ref{fig:mc}. This is a Monte Carlo (MC) simulation for 1000 datapoints for each of the four leak configurations, i.e., $leak=\{00,01,10,11\}$. This type of simulation is useful to determine how the oscillator will behave given that no silicon is identical and there is significant process variation involved. The mean frequencies of operation are 639, 671, 732, and 767MHz, which in turn correspond to the dynamic power values of 32.3, 34.2, 36.9, and 38.9uW depicted as distribution means in Fig.~\ref{fig:mc}.

When an adversary is devising his/her payload, many oscillator features can be configured. The number of bits being leaked at a time can be increased/decreased by using a MUX with more/less ports (or a chain of MUXes). The base frequency of oscillation can be increased/decreased by removing/adding more inverters to the common branch. The separation between the power consumption distributions can be tuned by uniquely sizing the BUFFs that drive the MUX ports. It is important that these distributions are distinguishable from one another, so the leaked data is clean and easy to read is. In Fig.\ref{fig:mc}, the minimal overlap between distributions happens in near-impossible scenarios where one BUFF and one BUFF only is disproportionately affected by process variation, which is an artefact of the over-pessimistic nature of MC simulation.

A single SPICE file with all the individual gate sizings is available in our repository. Yet, this is only one possible configuration, many more are possible by tuning the ring oscillator branches.

\section{Metal Stack and Routing Statistics}\label{app:metal}

\begin{figure}
\centering
\begin{tikzpicture}[scale=1]
    \begin{axis}[
            ybar stacked,
            nodes near coords,
            nodes near coords style={
                anchor=west,
                font=\tiny,
                /pgf/number format/fixed,
                xshift=0.1cm,
            },
            x tick label style={rotate=45, anchor=east,font=\footnotesize},
            y tick label style={font=\footnotesize},
            ylabel style={font=\footnotesize},
            symbolic x coords={Baseline, T64L16, T64L32, T64L64},
            enlarge x limits={abs=2.9*\pgfplotbarwidth},
            xtick style={draw=none},
            ylabel={Length ($\mu m$)},
            legend pos=outer north east,
            legend style={font=\footnotesize},
            width=0.85\columnwidth,
            ymin=0,
        ]
        \addplot [draw = black, fill=red!70, postaction={
        pattern=north east lines
        }] table[meta=metal,y=length]{\mtwotable};
        \addplot [draw = black, fill=green!80, postaction={
        pattern=north west lines
        }] table[meta=metal,y=length]{\mthreetable};
        \addplot [draw = black, fill=yellow, postaction={
        pattern=north east lines
        }] table[meta=metal,y=length]{\mfourtable};
        \addplot [draw = black, fill=red!20, postaction={
        pattern=north west lines
        }] table[meta=metal,y=length]{\mfivetable};
        \addplot [draw = black, fill=orange!50, postaction={
        pattern=north east lines
        }] table[meta=metal,y=length]{\msixtable};
        \addplot [draw = black, fill=wildstrawberry, postaction={
        pattern=north west lines
        }] table[meta=metal,y=length]{\mseventable};
        \legend{M2, M3, M4, M5, M6, M7}
    \end{axis}
\end{tikzpicture}
\vspace{-10pt}
\caption{Wirelength of the compromised designs versus the Trojan-free baseline.}
\label{fig:metal}
\end{figure}
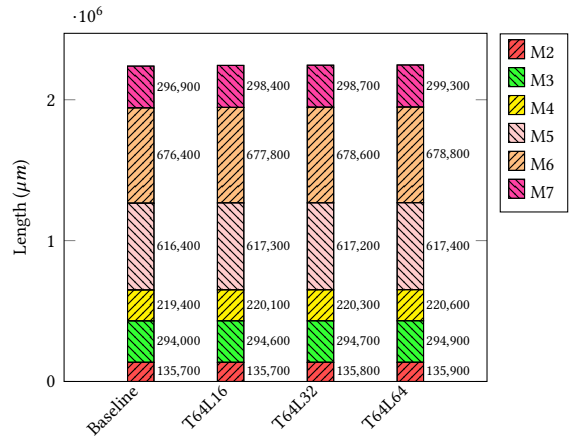

The considered metal stack in our layouts contains 9 metals. Metal 1 is reserved for intra-cell routing and cannot be used for signal routing. Metals 8 and 9 are reserved for the power grid and cannot be used for signal routing either. The remaining metals are 2-7. In Fig.~\ref{fig:metal}, we show the metal utilization of these metals (wirelength) for the baseline and three compromised versions of the speed-efficient $kali$ design depicted in Fig.~\ref{fig:kali_small_innovus}. We recall that the adversary considered is \texttt{A-FO} and the additional metal is incurred from the ECO operation. Notice how there are very small differences in metal utilization, which is an indication that the HTH insertion is stealthy and preserves the properties of the original design. This is the power of the ECO engine, here turned into a malicious attack vector.





\pgfplotstableread[row sep=\\,col sep=&]{
    circuit  & metal & length \\
    Baseline & M2    & 1.357e+05 \\
    T64L16   & M2    & 1.357e+05 \\
    T64L32   & M2    & 1.358e+05 \\
    T64L64   & M2    & 1.359e+05 \\
}\mtwotable

\pgfplotstableread[row sep=\\,col sep=&]{
    circuit  & metal & length \\
    Baseline & M3    & 2.940e+05 \\
    T64L16   & M3    & 2.946e+05 \\
    T64L32   & M3    & 2.947e+05 \\
    T64L64   & M3    & 2.949e+05 \\
}\mthreetable

\pgfplotstableread[row sep=\\,col sep=&]{
    circuit  & metal & length \\
    Baseline & M4    & 2.194e+05 \\
    T64L16   & M4    & 2.201e+05 \\
    T64L32   & M4    & 2.203e+05 \\
    T64L64   & M4    & 2.206e+05 \\
}\mfourtable

\pgfplotstableread[row sep=\\,col sep=&]{
    circuit  & metal & length \\
    Baseline & M5    & 6.164e+05 \\
    T64L16   & M5    & 6.173e+05 \\
    T64L32   & M5    & 6.172e+05 \\
    T64L64   & M5    & 6.174e+05 \\
}\mfivetable

\pgfplotstableread[row sep=\\,col sep=&]{
    circuit  & metal & length \\
    Baseline & M6    & 6.764e+05 \\
    T64L16   & M6    & 6.778e+05 \\
    T64L32   & M6    & 6.786e+05 \\
    T64L64   & M6    & 6.788e+05 \\
}\msixtable

\pgfplotstableread[row sep=\\,col sep=&]{
    circuit  & metal & length \\
    Baseline & M7    & 2.969e+05 \\
    T64L16   & M7    & 2.984e+05 \\
    T64L32   & M7    & 2.987e+05 \\
    T64L64   & M7    & 2.993e+05 \\
}\mseventable

\pgfplotstableread[row sep=\\,col sep=&]{
    metal & length    \\
    M2    & 1.357e+05 \\
    M3    & 2.940e+05 \\
    M4    & 2.194e+05 \\
    M5    & 6.164e+05 \\
    M6    & 6.764e+05 \\
    M7    & 2.969e+05 \\
    }\kalismall 

\pgfplotstableread[row sep=\\,col sep=&]{
    metal & length    \\
    M2    & 1.357e+05 \\
    M3    & 2.946e+05 \\
    M4    & 2.201e+05 \\
    M5    & 6.173e+05 \\
    M6    & 6.778e+05 \\
    M7    & 2.984e+05 \\
    }\kalismalltrojanLsixteen
    
\pgfplotstableread[row sep=\\,col sep=&]{
    metal & length    \\
    M2    & 1.358e+05 \\
    M3    & 2.947e+05 \\
    M4    & 2.203e+05 \\
    M5    & 6.172e+05 \\
    M6    & 6.786e+05 \\
    M7    & 2.987e+05 \\
    }\kalismalltrojanLthirtytwo

\pgfplotstableread[row sep=\\,col sep=&]{
    metal & length    \\
    M2    & 1.359e+05 \\
    M3    & 2.949e+05 \\
    M4    & 2.206e+05 \\
    M5    & 6.174e+05 \\
    M6    & 6.788e+05 \\
    M7    & 2.993e+05 \\
    }\kalismalltrojanLsixtyfour

\pgfplotstableread[row sep=\\,col sep=&]{
    metal & length    \\
    M2    & 1.649e+05 \\
    M3    & 3.444e+05 \\
    M4    & 2.852e+05 \\
    M5    & 5.188e+05 \\
    M6    & 4.474e+05 \\
    M7    & 1.500e+05 \\
    }\kalibig
    
\end{document}